\documentclass[11pt]{article}
\usepackage{graphicx} % Required for inserting images

\usepackage{fancyhdr}
\usepackage{isomath}
\usepackage{mathtools} % asmmath included here
\usepackage{amsbsy}
\usepackage{amssymb}
\usepackage{amscd,amsfonts}
\usepackage{bigints}
\usepackage{graphicx}
\usepackage{verbatim}
\usepackage{euscript}
\usepackage{alltt}
\usepackage{stmaryrd}
\usepackage{relsize}
\usepackage{enumerate}
\usepackage{url}
\usepackage[stable]{footmisc}
\usepackage{breakurl}
\usepackage{hyperref}
\usepackage{comment}
\usepackage{overpic}

\usepackage[maxbibnames=9]{biblatex}
\usepackage[font=small,labelfont=bf]{caption}
\usepackage[font=small,labelfont=bf]{subcaption}
\usepackage{float}
\usepackage{mwe}
\usepackage{xcolor}
\usepackage[super]{nth}
\usepackage{soul}

\DeclareGraphicsExtensions{.eps,.pdf}
\usepackage{amsmath}
\usepackage{amsbsy}
\usepackage{amssymb}
\usepackage{amscd}
\usepackage{amsfonts}
\usepackage{isomath}

\newcommand{\R}{\mathbb R}

\newcommand{\beq}{\begin{equation}}
\newcommand{\eeq}{\end{equation}}
\newcommand{\beqs}{\begin{eqnarray}}
\newcommand{\eeqs}{\end{eqnarray}}
\newcommand{\beql}{\begin{equation} \label}
\newcommand{\half}{\frac{1}{2}}

\newcommand{\calH}{{\cal H}}

\newcommand{\p}{\partial}
\newcommand{\dee}{\mathcal{D}}
\newcommand{\scl}{\mathcal{L}}

\usepackage[margin=1in]{geometry}
\usepackage{amsmath}
\usepackage{booktabs}

\date{}
\begin{document}
\title{Solving the Dissipation Inequality not as a constitutive restriction}

\author{Maximiliano Larrain Silva\thanks{Department of Civil \& Environmental Engineering, Carnegie Mellon University, Pittsburgh, PA 15213, email: maximill@andrew.cmu.edu.} $\qquad$ Amit Acharya\thanks{Department of Civil \& Environmental Engineering, and Center for Nonlinear Analysis, Carnegie Mellon University, Pittsburgh, PA 15213, email: acharyaamit@cmu.edu.}}

\maketitle
\begin{abstract}
\noindent 
A solution procedure is formulated and solved for treating  the nonlinear Dissipation Inequality as a constraint equation within continuum mechanics, and  allowing for incomplete knowledge of constitutive behavior. The scheme is demonstrated in the context of the rate-dependent, elastoplastic response of a bar, resulting in a nonlinear problem of constrained optimization. Both closed form and computational results are developed. The computational solutions utilize a sequence of convex optimization problems, and  are shown to be accurate. In the example considered, the approach is shown to automatically correct an (intentionally) faulty constitutive specification, resulting in the solution to be in accord with the fundamental postulates of continuum mechanics.

\end{abstract}

\section{Introduction}
In solving problems of continuum mechanics, the Second Law of thermodynamics or the Dissipation Inequality is difficult to accommodate as a constraint. It is also a fact that it is never possible to prescribe a model for material behavior with absolute certainty. In this communication, we study a simple model problem intentionally designed to embody these questions, and solve it using the ideas and scheme initiated in \cite{acharya2025second}. The model problem concerns the quasi-static response of a rate-dependent elastic-plastic bar whose specified constitutive equation is intentionally chosen to violate the  Dissipation Inequality over an interval of time, in the spirit of  \cite[Sec.~7.1]{needleman2023discrete}. We seek a self-consistent procedure that takes as input such a specification and nevertheless produces a solution that does not violate the Second Law, with `minimal' deviation from the specified constitutive response. An exact as well as a computational formulation is developed for the problem. It is shown how the ingredients of the proposed scheme attain its goals, allowing the solution of the problem in a well-set manner.
\section{A simple model problem}
\label{sec:sec2}
\begin{figure}
    \centering
    \includegraphics[width=0.5\linewidth]{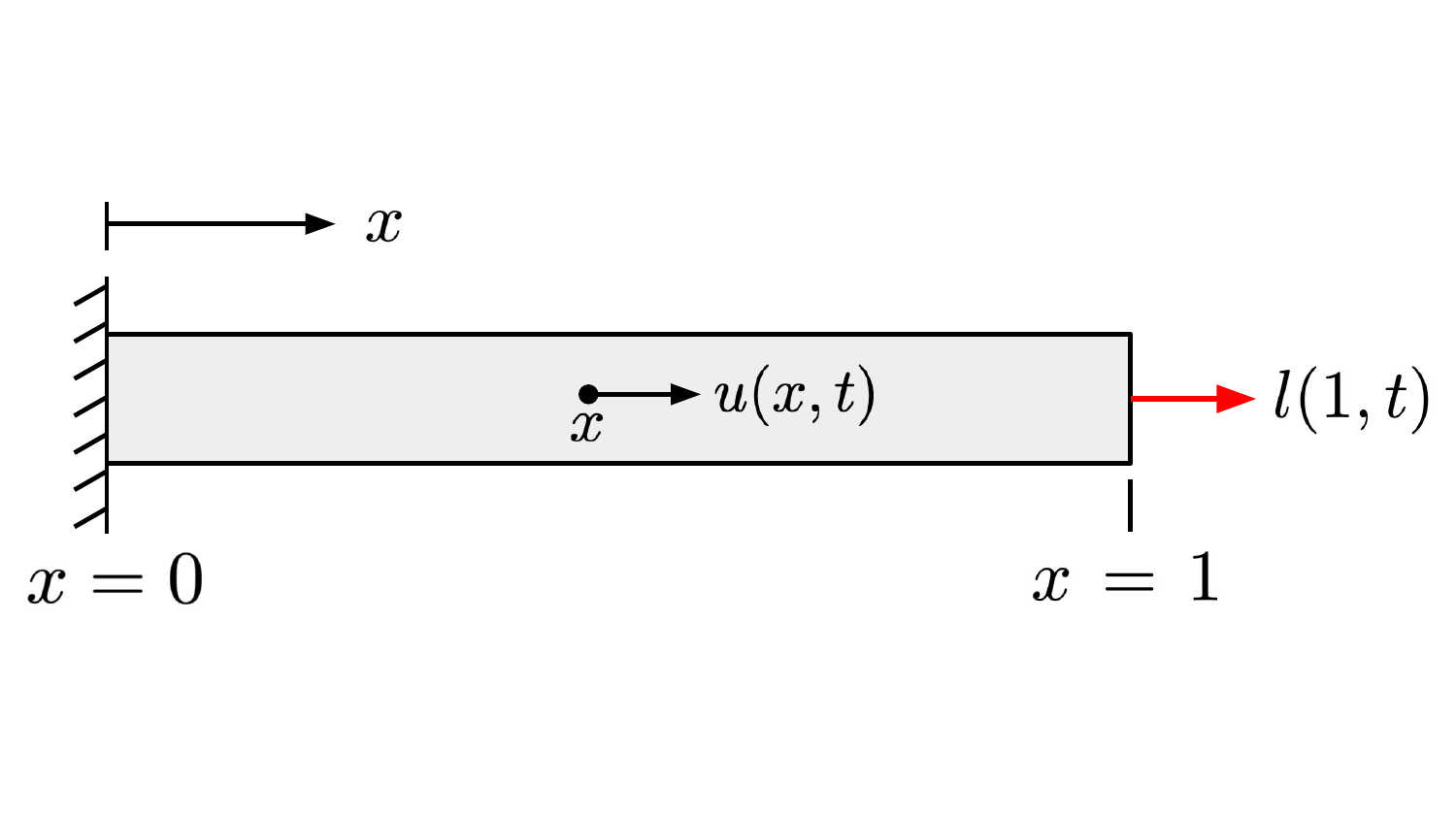}
    \caption{Schematic of the physical problem.}
    \label{fig:physical_set_up}
\end{figure}

Consider a 1-d bar made of an elastic-plastic material occupying the region $x \in [0,1]$ in the reference configuration, undergoing quasi-static stretching under a prescribed force (per unit cross-section area) $t \mapsto l(t)$ at the right end as a function of time. The left end is assumed to be fixed, i.e., $u(0,t) = 0$, where $(x,t) \mapsto u(x,t)$ is the displacement of the bar (see Fig. \ref{fig:physical_set_up}). With notation that subscript $t,x$ represent partial derivatives w.r.t.~$t,x$, respectively, if $(x,t) \mapsto \sigma(x,t)$ represents the axial stress in the bar, equilibrium 
\begin{equation}\label{eq:equil}
    \sigma_x = 0
\end{equation}
implies that the stress is only a function time in the bar given by
\begin{equation}\label{eq:eq_solved}
    \sigma(x, t) = \sigma(1,t) = l(t).
\end{equation}
%%Henceforth, we refer to $f(t)$ as $\sigma(t)$.
Let $\psi$ be the stored energy density for the material. Then the Second Law/Dissipation Inequality is given by
\begin{equation}\label{eq:sec_law}
    \sigma u_t \big|_{x = 1} - \sigma u_t \big|_{x = 0} - \psi_t \geq 0.
\end{equation}
Assuming a specified constitutive equation for stress and stored energy of the form
\begin{equation}\label{eq:const_eqn}
    \begin{aligned}
       \sigma = \sigma^c & = E (u_x - p) \\
        \psi^c & = \half E (u_x - p)^2
    \end{aligned}
\end{equation}
where $E$ is the Young's modulus, and $(x,t) \mapsto p(x,t)$ is the plastic strain in the bar. Using \eqref{eq:equil}, \eqref{eq:const_eqn} and the boundary condition on the displacement at the left end of the bar, \eqref{eq:sec_law} takes the form
\begin{equation}\label{eq:2_law_red}
    \sigma p_t \geq 0.
\end{equation}
Suppose the specified constitutive equation for the evolution of $p$ were to be given by
\begin{equation}\label{eq:const_p}
    p_t = f^c(\sigma,t).
\end{equation}
For simplicity let us assume that the loading $t \mapsto l(t)  \geq 0$. Given an initial condition $p(0) = p_0$ the plastic strain evolution is completely determined from solving \eqref{eq:const_p} in the form
\[
p_t = f^c(l(t), t); \qquad p(0) = p_0,
\]
and, therefore, if $f^c$ were to be an `arbitrarily' chosen physically motivated function, satisfaction of the Dissipation Inequality \eqref{eq:2_law_red}
\[
l(t)f^c(l(t),t) \geq 0
\]
cannot be guaranteed, unless the constitutive restriction
\begin{equation}\label{eq:const_restric}
    \sigma^c f^c \geq 0
\end{equation}
is imposed on the choice of the constitutive equation for the plastic strain rate.

We would now like to consider a formulation of the problem where the best possible physical choices of constitutive assumptions, i.e., estimates of material behavior, as deemed by a practitioner can be made and the Second Law satisfied without viewing it as a restriction on constitutive response as in \eqref{eq:const_restric}. Noting that a constitutive response can never be claimed to be known with certainty, we assume that the evolution of the plastic strain is given by
\begin{equation} \label{eq:constit_eq_fc_+_a}
p_t (x,t) = f^c(\sigma(x,t), t) + a(x,t)
\end{equation}
where the fields $(x,t) \mapsto (u(x,t), p(x,t), a(x,t), s(x,t))$ are to be determined by solving \eqref{eq:equil}-\eqref{eq:2_law_red} in the form
\begin{equation}\label{eq:gov_eqn}
    \begin{aligned}
        l \big(f^c(l,t) + a \big) - \half s^2 & = 0\\
        p_t - f^c(l,t) - a & = 0\\
        p(0) & = p_0,
    \end{aligned}
\end{equation}
where $p_0$ is a specified initial condition on the plastic strain.

The function $(1/2) s^2$ represents the dissipation density, and in this formulation, $s$ is a field we solve for as well. Given \eqref{eq:equil}, it suffices to solve the problem with $x$-independent fields, which is an ansatz we adopt. Once a solution for $t \mapsto p(t)$ is obtained, \eqref{eq:const_eqn} is solved in the form $u_x(x,t) = E^{-1} l(t) + p(t)$ and integrated w.r.t. $x$ with b.c.~$u(0,t) = 0$ to obtain the displacement of the bar.

Thus the problem becomes a question of obtaining, in a well-set manner, the functions $t \mapsto (p(t), s(t), a(t))$ which satisfy \eqref{eq:gov_eqn}. It is clear that we have two equations in three functions and, in general, this is going to be a practical impediment even if conceptually the non-uniqueness is accepted.

To have a physically reasonable and practically tractable problem, we follow \cite{acharya2025second} and require that $a$ be as small as possible, reflecting faith in the constitutive specification deployed, but allowing, at the same time, for non-negative dissipation and the impossibility of knowing material response with infinite precision. Since we have employed a constitutive response function $f^c$ and would like the dissipation in the system to be dictated solely by that modeling knowledge to the extent possible without violating the fundamental constraints embodied in \eqref{eq:gov_eqn}, we also require $(1/2)s^2$ to be as small as possible without violating the constraints \eqref{eq:gov_eqn}.

Thus, we consider the minimization problem on the time interval $[0,T]$ given by
\begin{equation}\label{eq:min}
\min_{a,s} \int_0^T \calH(a,s) \,dt; \qquad \calH(a,s) := \half c_a a^2 + \half c_s s^2
\end{equation}
subject to the constraints \eqref{eq:gov_eqn} written in the form
\begin{subequations}\label{eq:a_s_p}
    \begin{align}
        a(s) & = \frac{1}{2l}s^2 - f^c \label{eq:a(s)}\\
        p_t & = \frac{1}{2 l} s^2; \qquad p(0) = p_0. \label{eq:p_evol}
    \end{align}
\end{subequations}
Noticing the form of the constrained minimization problem, the optimization can be considered pointwise in time without any differential constraints for the objective
\[
h(s) = \half c_a (a(s))^2 + \half c_s s^2,
\]
with $a(s)$ given by \eqref{eq:a(s)}, and $c_s$ and $c_a$ being positive material constants with the physical dimension of the ratio $c_s/c_a$ being $(Time.stress)^{-1}$.

It is easily checked that the minimum is attained at 
\begin{equation}\label{eq:s2_analytic}
    s^2(t) = \begin{cases} 
     0 \qquad \mbox{for} \qquad f^c(l(t),t) - \frac{c_s}{c_a} l(t) < 0\\
     \\
     2l(t) \left(f^c(l(t),t) - \frac{c_s}{c_a} l(t) \right) \qquad \mbox{for} \qquad f^c(l(t),t) - \frac{c_s}{c_a} l(t) \geq 0,
     \end{cases}
\end{equation}
with \eqref{eq:p_evol} evolved according to the above specification of $s^2$.

For $\sigma_0, \hat{\gamma}$ being typical stress and strain rate scales for the problem entering through the specified constitutive equation $f^c$, e.g., the yield stress and reference plastic strain rate, it is interesting to note the solution of the problem in the limit
\begin{equation*}
    \frac{c_s \sigma_0}{c_a \hat{\gamma}} \to 0
\end{equation*}
is given by
\begin{equation}
s^2(t) = \begin{cases}\label{eq:s2_limit}
0 \quad \mbox{for} \quad f^c(l(t),t) < 0 \\
\\
2 l(t) \, f^c(l(t),t) \quad \mbox{for} \quad f^c(l(t),t) \geq 0
\end{cases};
\qquad
p_t = \frac{s^2}{2 l} = 
\begin{cases}
    0 \quad \mbox{for} \quad f^c < 0 \\
    f^c \quad \mbox{for} \quad f^c \geq 0. 
\end{cases}
\end{equation}
This is the naturally expected result selected by our scheme out of the infinite family of solutions to the system \eqref{eq:gov_eqn} consisting of force equilibrium, constitutive equation, and the Second Law. The infinite family is parametrized by any choice of the dissipation function $t \mapsto (1/2) s^2(t)$, as evident from \eqref{eq:a_s_p}. The approach above selects a solution of this family. 

In what follows we evaluate a computational scheme, originally designed for solving systems of partial differential equations (which obviously includes ordinary differential and algebraic equations) using methods of the calculus of variations \cite{ka1,ka2,kpa,sga,suku_ach,AV_nash}. It tries to obtain the desired solution obtained above of the constrained minimization problem in a stable fashion without exploiting special simplifying features, e.g., not eliminating differential constraints as done above.

\section{Approximation based on a convex dual variational principle}
This section presents the variational dual formulation of the governing equations \eqref{eq:min}-\eqref{eq:a_s_p}, following the general theory proposed in \cite{ach3}, \cite[Sec.~3]{AV_nash}.

We use the notation $U:=(p,s,a)$ denoting the primal fields and $D:=(\alpha,\beta)$ the dual fields. A pre-dual functional $\widehat{S}_H[U,D]$ is formed by taking the scalar product of equations \eqref{eq:a_s_p} with the dual fields $D$, integrating by parts, substituting the prescribed initial condition for $p$ (ignoring for now the final-time boundary condition that arises on integration by parts but is not specified as part of the primal physical problem \eqref{eq:a_s_p}) and subtracting an auxiliary potential $H$, resulting in
\begin{equation}\label{eq:predual2}
        \widehat{S}_H[U,D]=\int_0^T \Big( l \big(f^c + a \big) - \frac{1}{2}s^2\Big)\alpha -p\beta_t - f^c\beta - a\beta - H(U,\bar{U}) \,dt \ - \ p_0\beta(0).
\end{equation}
In the above, $\bar{U}:=(\bar{p},\bar{s},\bar{a})$ are prescribed functions of time called `base states' and $H(U,\bar{U})$ is specified as
\begin{equation}\label{eq:potential_H}
    H(U,\bar{U})=\frac{c_p}{2}(p-\bar{p})^2+\frac{c_s}{2}(s-\bar{s})^2+\frac{c_a}{2}(a-\bar{a})^2; \qquad c_p, c_s, c_a > 0.
\end{equation}
A part of the auxiliary potential $H$ is chosen to resemble the physical objective function $\mathcal{H}$ \eqref{eq:min} from Section \ref{sec:sec2}; the weights $c_s$ and $c_a$ in $\mathcal{H}$ match those in $H$. This is natural and aids in the selection of the solution representing  a minimal control $a$ and dissipation $s^2/2$. This design of the auxiliary function $H$ is slightly different from the case when the question is solely to solve a system of equations without involving a physical functional/objective to be optimized, as in  \cite{ka1,ka2,sga,kpa,AV_nash}. In that case, the design of the auxiliary function is entirely governed by considerations of mathematical convenience for obtaining a solution to the system.

For simplicity, we refer to the composition $t \mapsto f^c \circ l(t)$ as $t \mapsto f^c(t)$. In this specific problem, $f^c$ becomes a specified function of time due to the simplification \eqref{eq:eq_solved}. 

Define $\mathcal{D}:=(D,D_t)$ and the Lagrangian $\mathcal{L}_H(U,\mathcal{D},\bar{U})$ as
\begin{equation}
    \mathcal{L}_{H}(U,\mathcal{D},t)=\Big( l \big(f^c + a \big) - \frac{1}{2}s^2\Big)\alpha -p\beta_{t} - f^c\beta - a\beta - H(U,\bar{U}).
\end{equation}
Then, the `dual-to-primal' (DtP) map, denoted $U^{(H)}(\mathcal{D},\bar{U},t)$, is obtained by solving $\p_U \scl_H(U, \dee, \bar{U}, t) = 0$ for $U$ as $(\dee, \bar{U}, t) \mapsto U^{(H)}(\dee, \bar{U},t)$ satisfying
\begin{equation}\label{eq:dLhdu_0}
    \frac{\partial \mathcal{L}_H}{\partial U}\left(U^{(H)}(\mathcal{D},\bar{U}),\mathcal{D},t \right)=0. 
\end{equation}
In this particular case, $U^{(H)}(\mathcal{D},\bar{U}, t)$ is obtained as
\begin{equation}\label{eq:dtp_map}
    \begin{aligned}
      p_H(\mathcal{D},\bar{U},t)& =\bar{p}-\frac{\beta_t}{c_p}\\
      s_H(\mathcal{D},\bar{U},t)& =\frac{c_s\bar{s}}{\alpha+c_s}\\
      a_H(\mathcal{D},\bar{U},t)& =\bar{a}+\frac{l\alpha -\beta}{c_a}.
    \end{aligned}
\end{equation}
The \emph{dual} functional $S_H[D]$ is then obtained by substituting the DtP map \eqref{eq:dtp_map} into the pre-dual functional \eqref{eq:predual2} as
\begin{equation}\label{eq:sh}
    \begin{aligned}
        S_H[D]& = \widehat{S}_H \left[U^{(H)}(\mathcal{D},\bar{U},t), D \right]=\int_0^T \mathcal{L}_H\Bigl(U^{(H)}(\mathcal{D},\bar{U},t),\mathcal{D},t\Bigr)\hspace{2mm}dt \hspace{2mm}-p_0\beta(0)\\
        & =\int_0^T \Big( l \big(f^c + a_H \big) - \frac{1}{2}s_H^2\Big)\alpha -p_H\beta_t - f^c\beta - a_H\beta - H(U^{(H)},\bar{U})\hspace{2mm}dt\hspace{2mm}-p_0\beta(0).\\
    \end{aligned}
\end{equation}
Its first variation about a state $D$ in the direction $\delta D$ is given by
\begin{multline}\label{eq:deltaSH}
    \delta S_H\Big|_{\delta D}[D;\bar{U}]=\int_0^T\frac{\partial \mathcal{L}_H}{\partial \mathcal{D}}\left(U^{(H)}(\mathcal{D},t),\mathcal{D},t\right)\cdot\delta  \mathcal{D}\hspace{2mm}dt\hspace{2mm}-p_0\delta \beta(0)\\
    =\int_0^T \Big(l \big(f^c + a_H \big) - \frac{1}{2} s_H^2\Big)\delta \alpha  -p_H\delta\beta_t  - f^c\delta \beta - a_H\delta \beta \hspace{2mm}dt\hspace{2mm}-p_0\delta \beta(0).
\end{multline}
Since $\mathcal{L}_H$ is affine in its argument $\mathcal{D}$, and using \eqref{eq:dLhdu_0} as well as the fact that $\beta(t)$ has a Dirichlet boundary condition specified at $t=T$ so that $\delta \beta(T) = 0$, the Euler-Lagrange equations and natural boundary conditions of the dual functional $S_H$ are exactly the primal equations shown in \eqref{eq:a_s_p} with the primal variables $U$ substituted by the DtP map, $U^{(H)}$. Indeed, in this specific case we have
\begin{equation}
        \delta S_H\Big|_{\delta D}[D] =\int_0^T \Big(l \big(f^c + a_H \big) - \half s_H^2\Big)\delta \alpha + \Big((p_{H})_t - f^c - a_H\Big)\delta \beta \hspace{2mm}dt+\big(p_H(0)-p_0\big)\delta \beta(0),
\end{equation}
from which the E-L equations and the initial condition can be read off.

An important feature of the formulation which makes the dual functional \emph{convex}, with its associated variational principle a minimum (as opposed to an extremum) principle, is ensured by requiring that the Dual functional $S_H$ is restricted to the domain in $D$-space for which 
\begin{equation}\label{eq:DtP}
    -\partial_{UU}\mathcal{L}_H\left(\hat{U}(t),\mathcal{D}(t),\bar{U}(t),t\right)>0 \quad \mbox{ for each } t \in (0,T); \qquad \hat{U}(t) := U^{(H)}\left(\dee(t), \bar{U}(t),t\right),
\end{equation}
see \cite[Sec.~3]{AG_control}, \cite[Sec.~3] {AV_nash}. We employ this restriction in all that follows by utilizing appropriate changes of base states. We refer to this restricted domain of the dual functional $S_H$ as the DtP zone.

In this example, the Hessian of the Lagrangian, $\p_{UU} \scl_H (U,\dee, \bar{U}, t)$, is diagonal, so the condition reduces to requiring that each diagonal entry (i.e., each eigenvalue) be strictly negative:

\begin{equation}
    -\frac{\partial^2 }{\partial p^2}\mathcal{L}_H=c_p > 0 ; \qquad -\frac{\partial^2}{\partial s^2} \mathcal{L}_H=\alpha+c_s > 0; \qquad -\frac{\partial^2 }{\partial a^2}\mathcal{L}_H=c_a > 0 .
\end{equation}
Since $c_p$, $c_a$ $>0$, the only non trivial restriction defining the DtP zone is
\begin{equation}\label{eq:dtp_zone}
    \alpha(t)>-c_s \hspace{10mm} \forall t \in (0,T).
\end{equation}

Having derived the dual functional $S_H[D]$ and its first variation, our goal is to find a stationary point of $S_H[D]$. This is achieved in two consecutive phases: first, the gradient flow scheme presented in \cite[Sec.~3]{AV_nash}
drives the initial approach to the solution; second, this solution is refined using the Newton Raphson scheme with step size control presented in \cite{kpa}. Here, the Newton Raphson phase is modified to also include a restriction on the $L^2$ norm of the gradient of $S_H$, which must not increase throughout this phase, since $S_H$ is convex with the restriction of its domain discussed surrounding \eqref{eq:DtP}. In both phases, the finite element method is used to discretize and approximately solve the problem.

For the gradient flow, we follow \cite[Section 3.2]{AV_nash}, and think of a (fake, time-like) variable $\mathsf{s}$ along which a gradient descent of the functional $S_{H_k}$ is executed. Here, $H_k$ represents the auxiliary potential used in stage $k$ of the gradient flow, parametrized by the base state $\bar{U^k}$, which is held fixed throughout that stage. Each stage comprises the interval $[0,\tilde{\mathsf{s}}^*_k)$, with $\tilde{\mathsf{s}}^*_k$ being the (fake)time instant at which the gradient flow achieves convergence, or one of the stopping criteria for stage $k$ is met (see \tableautorefname{ \ref{tab:algorithm}}). We consider dual unknown fields defined by $(t,\mathsf{s}) \mapsto D^k(t,\mathsf{s})$, and corresponding variations $\delta D$ to execute the following gradient flow:
\begin{equation}\label{eq:grad_flow}
    \int_0^T \delta D(t)\cdot \frac{\partial D^k}{\partial \mathsf{s}}(t,\mathsf{s})dt=-\int_0^T\frac{\delta S_{H_k}}{\delta D}[D^k;\bar{U}^k](t,\mathsf{s})\cdot \delta D(t)\,dt.
\end{equation}
Let $\Omega = \{t: t \in (0,T)\}$ denote the physical time domain, which we discretize using a uniform finite element mesh. The trial and test functions are chosen to be piecewise linear and globally continuous, dependent solely on $t$. These shape functions are denoted by $N^{(\cdot)}$, where $(\cdot)$ denotes the index of the node associated with that function.
Using this discretization, the dual fields and their respective variations are approximated by the expressions shown in \eqref{eq:dual_fem_aprox} and \eqref{eq:deltadual_fem_aprox}, where the summation convention is used. Throughout, the uppercase index $B$ denotes the node associated with the trial function, while the uppercase index $A$ denotes the node associated with the test (variation) function. 
\begin{equation}\label{eq:dual_fem_aprox}
    \alpha(t,\mathsf{s})=\alpha^{B}(\mathsf{s})N^{B}(t)   \hspace{10mm}\beta(t,\mathsf{s})=\beta^{B}(\mathsf{s})N^{B}(t) 
\end{equation}
\begin{equation}\label{eq:deltadual_fem_aprox}
    \delta \alpha(t)=\delta \alpha^{A}N^{A}(t)   \hspace{10mm}\delta \beta(t)=\delta \beta^{A}N^{A}(t).
\end{equation}
As shown in \eqref{eq:dual_fem_aprox}, the dual fields $\alpha(t,\mathsf{s})$ and $\beta(t,\mathsf{s})$ depend on the variable $\mathsf{s}$ only through the coefficients that multiply the shape functions, not through the shape functions themselves (which depend only on physical time $t$). Expanding \eqref{eq:grad_flow} using \eqref{eq:deltaSH}, we have
\begin{multline}\label{eq:gf_eqs_1}
    \int_0^T \delta \alpha(t)\frac{\partial \alpha}{\partial \mathsf{s}}(t,\mathsf{s})dt +
    \int_0^T \delta \beta(t)\frac{\partial \beta}{\partial \mathsf{s}}(t,\mathsf{s})dt=\\
     -\int_0^T \Big(l \big(f^c + a_H \big) - \frac{1}{2} s_H^2\Big)\delta \alpha(t)  -p_H\delta\beta_t(t)  - f^c\delta \beta(t) - a_H\delta \beta(t) \hspace{2mm}dt\hspace{2mm}-p_0\delta \beta(0).
\end{multline}
Note that $\partial/\partial \mathsf{s}$ denotes differentiation with respect to the fictitious gradient flow time, whereas the subscript $t$ appearing in $\delta \beta_t$ denotes differentiation with respect to physical time, consistent with \eqref{eq:deltaSH}. Substituting \eqref{eq:dual_fem_aprox} and \eqref{eq:deltadual_fem_aprox} in \eqref{eq:gf_eqs_1}, we obtain
\begin{multline}\label{eq:gf_eqs_2}
\delta \alpha^{A}\Big(\int_0^T N^{A}(t) \frac{\partial \alpha^B}{\partial \mathsf{s}}(\mathsf{s})N^B(t)dt\Big) +
    \delta \beta^{A}\Big(\int_0^T N^{A}(t) \frac{\partial \beta^B}{\partial \mathsf{s}}(\mathsf{s})N^B(t)dt\Big)=\\
    -\delta \alpha^{A}\Big(\int_0^T \Big(l \big(f^c + a_H \big) - \frac{1}{2} s_H^2\Big)N^{A}(t)dt\Big)   -\delta \beta^{A}\Big(\int_0^T-p_HN_t^{A}(t) \\ - f^cN^{A}(t) - a_HN^{A}(t) dt\hspace{2mm}-p_0N^{A}(0)\Big).
\end{multline}
Factoring out $\delta \alpha^{A}$ and $\delta \beta^{A}$, \eqref{eq:gf_eqs_2} becomes:
\begin{equation}\label{eq:gf_eqs_3}
    \delta \alpha^{A}\Big(M^{AB}\frac{\partial \alpha^B}{\partial \mathsf{s}}(\mathsf{s})+R_{(\alpha)}^A\Big)+\delta \beta^{A}\Big(M^{AB}\frac{\partial \beta^B}{\partial \mathsf{s}}(\mathsf{s})+R_{(\beta)}^A\Big)=0,
\end{equation}
where the mass matrix $M^{AB}$ and the residual vectors $R_{(\alpha)}^A$ and $R_{(\beta)}^A$ are defined as

\begin{equation}\label{eq:MassMatrix}
    M^{AB}:=\int_0^T N^{A}(t) N^B(t) \ dt,
\end{equation}
\begin{equation}\label{eq:resid_alpha}
    R_{(\alpha)}^A(t):=\int_0^T \Big(l \big(f^c + a_H \big) - \frac{1}{2} s_H^2\Big)N^{A}(t) \ dt,
\end{equation}
\begin{equation}\label{eq:resid_beta}
    R_{(\beta)}^A(t):=\int_0^T-p_HN_t^{A}(t) - f^cN^{A}(t) - a_HN^{A}(t) \ dt-p_0N^{A}(0).
\end{equation}
Since \eqref{eq:gf_eqs_3} must hold for all $\delta \alpha^{A}$ and $\delta \beta^{A}$, we obtain the following system of equations:

\begin{equation}\label{eq:gf_eqs_4}
     M^{AB}\frac{\partial \alpha^B}{\partial \mathsf{s}}(\mathsf{s})=- R_{(\alpha)}^A; \qquad M^{AB}\frac{\partial \beta^B}{\partial \mathsf{s}}(\mathsf{s})=-R_{(\beta)}^A \hspace{10mm}\forall A.
\end{equation}
Finally, to integrate the system \eqref{eq:gf_eqs_4} in the fictitious time $\mathsf{s}$, we partition the interval $[0,\tilde{\mathsf{s}}^*_k)$ into discrete instants $\mathsf{s}_0:=0<\mathsf{s}_1<\mathsf{s}_2<\ldots$, with a step size $\Delta \mathsf{s}_n$ (not necessarily constant and determined by satisfying both \eqref{eq:dtp_zone} and having the $L^2$ norm of the residual non increasing from instant $\mathsf{s}_n$ to $\mathsf{s}_{n+1}$ (see \tableautorefname{ \ref{tab:algorithm}})), and denote by $\alpha^B(\mathsf{s}_n)$, $\beta^B(\mathsf{s}_n)$ the nodal coefficients of the dual functions at time $\mathsf{s}_{n}$. Approximating the derivatives of $\alpha^B$ and $\beta^B$ with respect to $\mathsf{s}$ using a forward difference scheme (forward Euler), and evaluating the residual $R_{(\alpha)}^A$ and $R_{(\beta)}^A$ using the known values $\alpha^B(\mathsf{s}_n)$, $\beta^B(\mathsf{s}_n)$ (through the DtP map $p_H$, $s_H$ and $a_H$), we obtain the linear system of equations shown in \eqref{eq:G_F_discr_eqn} to be solved to obtain $\alpha^B(\mathsf{s}_{n+1})$, $\beta^B(\mathsf{s}_{n+1})$, thus advancing from time $\mathsf{s}_n$ to $\mathsf{s}_{n+1}$:

\begin{equation}\label{eq:G_F_discr_eqn}
         M^{AB}\left( \frac{\alpha^B(\mathsf{s}_{n+1})-\alpha^B(\mathsf{s}_{n})}{\Delta \mathsf{s}_n}\right)=-R_{(\alpha)}^A \hspace{10mm}
         M^{AB}\left(\frac{\beta^B(\mathsf{s}_{n+1})-\beta^B(\mathsf{s}_{n})}{\Delta \mathsf{s}_n}\right)=-R_{(\beta)}^A\hspace{10mm}\forall A.
\end{equation}
For the second phase consisting of the Newton-Raphson scheme with step size control, we follow the algorithm presented in \cite{kpa}. The system Jacobian is computed by taking the variation of the residual \eqref{eq:deltaSH}, given by 
\begin{multline}\label{eq:Jacob1}
     J\Big|_{\delta D,d D}[D] =\int_0^T \Big( \big(l\frac{\partial a_H}{\partial \alpha} - s_H\frac{\partial s_H}{\partial \alpha}\big)d\alpha + l\frac{\partial a_H}{\partial \beta}d\beta \Big)\delta \alpha \, dt\\  +\int_0^T\Big(-\frac{\partial a_H}{\partial \alpha}d\alpha \delta \beta-\frac{\partial p_H}{\partial \beta_t}d\beta_t \delta\beta_t -\frac{\partial a_H}{\partial \beta}d\beta\delta \beta \Big) \, dt.
\end{multline}
Recalling the approximations for $\delta \alpha$ and $\delta \beta$ given in \eqref{eq:deltadual_fem_aprox} and introducing analogous approximations for the linearization directions $d\alpha$ and $d\beta$
\begin{equation}\label{eq:ddual_fem_aprox}
    d\alpha(t)=d\alpha^{C}N^{C}(t),   \hspace{10mm}d\beta(t)=d\beta^{D}N^{D}(t),
\end{equation}
the Jacobian in \eqref{eq:Jacob1} can be expressed as
\begin{multline}\label{eq:Jacob2}
     J\Big|_{\delta D,d D}[D] = \\
     \delta \alpha^{A} \ \left[ \int_0^T \big(l\frac{\partial a_H}{\partial \alpha} - s_H\frac{\partial s_H}{\partial \alpha}\big)N^{A}(t)N^{C}(t)\, dt \right] \ d\alpha^{C} \ + \ \delta \alpha^{A} \ \left[ \int_0^T l\frac{\partial a_H}{\partial \beta}N^{A}(t)N^{D}(t) \, dt \right] \ d\beta^{D} \\  
     + \ \delta \beta^{A} \ \left[\int_0^T-\frac{\partial a_H}{\partial \alpha}N^{A}(t) N^C(t)\, dt\right] \ d\alpha^{C} \ + \ \delta \beta^{A} \ \left[\int_0^T-\frac{\partial p_H}{\partial \beta_t}N^{A}_t(t) N^{D}_t(t) -\frac{\partial a_H}{\partial \beta}N^{A}(t) N^{D}(t) \, dt\right] \ d\beta^{D}.
\end{multline}
Denoting the main blocks of the system Jacobian as

\begin{equation}
    J_{(\alpha \alpha)}^{AC}=\int_0^T \big(l\frac{\partial a_H}{\partial \alpha} - s_H\frac{\partial s_H}{\partial \alpha}\big)N^{A}(t)N^{C}(t)dt; \qquad J_{(\alpha \beta)}^{AD}=\int_0^T  l\frac{\partial a_H}{\partial \beta}N^{A}(t)N^{D}(t)  dt\\ \nonumber
\end{equation}
\begin{equation}
    J_{(\beta \alpha)}^{AC}=\int_0^T  -\frac{\partial a_H}{\partial \alpha}N^{A}(t) N^C(t)dt; \qquad J_{(\beta \beta)}^{AD}=\int_0^T-\frac{\partial p_H}{\partial \beta_t}N^{A}_t(t) N^{D}_t(t) +\frac{\partial a_H}{\partial \beta}N^{A}(t) N^{D}(t)dt,
\end{equation}
the Newton-Raphson system, with unknown coefficients $d\alpha^{C}$ and $d\beta^{D}$, is given by
\begin{equation}\label{eq:NR_system}
    \begin{pmatrix} J_{(\alpha \alpha)}^{AC} & J_{(\alpha \beta)}^{AD} \\ &\\J_{(\beta \alpha)}^{AC} & J_{(\beta \beta)}^{AD} \end{pmatrix}
    \begin{pmatrix} d\alpha^{C} \\ \\d\beta^{D} \end{pmatrix}
    =
    -\begin{pmatrix} R_{(\alpha)}^A \\ \\R_{(\beta)}^A \end{pmatrix},
\end{equation}
where $R_{(\alpha)}^A$ and $R_{(\beta)}^A$ are defined by \eqref{eq:resid_alpha} and \eqref{eq:resid_beta}, respectively. 
Finally, after each Newton-Raphson iteration, the dual unknown coefficients are updated according to
\begin{equation}
    \begin{aligned}
        \alpha^C(\mathsf{s}_{n+1})=\alpha^C(\mathsf{s}_{n})+{\Delta \mathrm{\mathsf{s}}_n}d\alpha^C \\  
        \beta^D(\mathsf{s}_{n+1})=\beta^D(\mathsf{s}_{n})+{\Delta \mathrm{\mathsf{s}}_n}d\beta^D,
    \end{aligned}
\end{equation}
where $\Delta \mathsf{s}_n$ is the step size determined (as in the gradient flow phase) by satisfying the DtP zone condition \eqref{eq:dtp_zone} and having a non increasing $L^2$ norm of the residual moving from instant $\mathsf{s}_{n}$ to $\mathsf{s}_{n+1}$. The implementation details of the computational scheme are shown in the algorithm presented in Table \ref{tab:algorithm}.

\begin{table}%%[h]
\centering
\begin{tabular}{p{0.95\textwidth}}
\toprule
\textbf{Algorithm: Gradient Flow \& Newton-Raphson Iterations} \\
\midrule
% --- INITIALIZATION ---
\textbf{Initialization:} Set $c_p$, $c_s$, $c_a$, $tol_{DtP}$, $tol_{NR}$, $tol$, $\Delta \mathsf{s}_{min}$, $\Delta \mathsf{s}_{init}$ \\
\hspace{4mm} Set initial base states ($k=1$): $\bar{p}_1=\bar{P_0}$, $\bar{s}_1=\bar{S_0}$, $\bar{a}_1=\bar{A_0}$ \\
\hspace{4mm} Set $N_{elem}$, assemble mass matrix $M$ \\[6pt]
% --- STAGE ---
\textbf{1) $k_{\text{th}}$ stage:} \\[2pt]
\hspace{4mm} \textbf{1a.} Set $\mathsf{s}_0=0$ \\[2pt]
\hspace{4mm} \textbf{1b.} Set $\alpha_k^B(\mathsf{s}_0)=0$ and $\beta_k^B(\mathsf{s}_0)=0$ $\forall B$ (i.e., $D^{k}(\mathsf{s}_0)=0$) \\[2pt]
\hspace{4mm} \textbf{1c.} Set $\bar{p}_k^B(\mathsf{s}_0)=p^B_{H_{k-1}}(\tilde{\mathsf{s}}^*_{k-1})$, $\bar{s}_k^B(\mathsf{s}_0)=s^B_{H_{k-1}}(\tilde{\mathsf{s}}^*_{k-1})$ and $\bar{a}_k^B(\mathsf{s}_0)=a^B_{H_{k-1}}(\tilde{\mathsf{s}}^*_{k-1})$ $\forall B$ \\
\hspace{10mm} (i.e., $\bar{U}^k(\mathsf{s}_0) = U^{H_{k-1}}(\tilde{\mathsf{s}}^*_{k-1})$) \\[2pt]
\hspace{4mm} \textbf{1d.} Set $\Delta_{\mathsf{s}_{k}}=\Delta \mathsf{s}_{init}$ \\[2pt]
\hspace{4mm} \textbf{1e.} Solve DtP$_{map}$ using (1b) and (1c) to obtain $p_{H_k}(\mathsf{s}_0)$, $s_{H_k}(\mathsf{s}_0)$ and $a_{H_k}(\mathsf{s}_0)$ \\[2pt]
\hspace{4mm} \textbf{1f.} Compute residual \eqref{eq:resid_alpha}-\eqref{eq:resid_beta}, evaluate its $L_2$ norm, define it as $|rhs|_k(\mathsf{s}_0)$ \\[6pt]
% --- ITERATIONS ---
\textbf{2) Gradient Flow/Newton-Raphson Iterations:} \\[2pt]
\hspace{4mm} \textbf{2a.} \textbf{For} $n \geq 0$: \\[2pt]
\hspace{8mm} \textbf{i.} \textbf{if} $|rhs|_k(\mathsf{s}_n) \geq tol_{NR}$: \\
\hspace{16mm} Gradient Flow: Solve Eq.~\ref{eq:G_F_discr_eqn} $\rightarrow$ dual guess: $\alpha_*^B(\mathsf{s}_{n+1})$, $\beta_*^B(\mathsf{s}_{n+1})$ \\
\hspace{8mm} \textbf{else}: Newton-Raphson: Solve Eq.~\ref{eq:NR_system} $\rightarrow$ dual guess: $\alpha_*^B(\mathsf{s}_{n+1})$, $\beta_*^B(\mathsf{s}_{n+1})$ \\[2pt]
\hspace{8mm} \textbf{ii.} Solve DtP$_{map}$ using $\alpha_*^B(\mathsf{s}_{n+1})$, $\beta_*^B(\mathsf{s}_{n+1})$ $\rightarrow$ obtain $p_{H_k}(\mathsf{s}_{n+1})$, $s_{H_k}(\mathsf{s}_{n+1})$, $a_{H_k}(\mathsf{s}_{n+1})$ \\[2pt]
\hspace{8mm} \textbf{iii.} \textbf{if} $a_{H_k}(\mathsf{s}_{n+1}) > -c_a + tol_{DtP}$ \textbf{(DtP zone check)}: \\
\hspace{16mm} Compute $|rhs|_k(\mathsf{s}_{n+1})$ \\
\hspace{16mm} \textbf{if} $|rhs|_k(\mathsf{s}_{n+1}) \leq |rhs|_k(\mathsf{s}_{n})$ \textbf{(descent condition check)}: \\
\hspace{24mm} Accept and save dual guess: $\alpha^B(\mathsf{s}_{n+1})=\alpha_*^B(\mathsf{s}_{n+1})$, $\beta^B(\mathsf{s}_{n+1})=\beta_*^B(\mathsf{s}_{n+1})$ \\
\hspace{24mm} Save $|rhs|_k(\mathsf{s}_{n+1})$ \\
\hspace{24mm} Save time: $\mathsf{s}_{n+1}=\mathsf{s}_n+\Delta \mathsf{s}_{n}$ \\
\hspace{24mm} \textbf{if} $|rhs|_k(s_{n+1}) \leq tol$: \\
\hspace{32mm} \textbf{return} $p_{H_k}(\mathsf{s}_{n+1})$, $s_{H_k}(\mathsf{s}_{n+1})$, $a_{H_k}(\mathsf{s}_{n+1})$ — \textbf{Convergence Achieved} \\
\hspace{24mm} \textbf{else}: Set $\Delta \mathsf{s}_{n+1}=\Delta \mathsf{s}_n$, $n=n+1$, go to \textbf{2a.i} \\[2pt]
\hspace{16mm} \textbf{else} (descent fails): Set $\Delta \mathsf{s}_{n}=\Delta \mathsf{s}_{n}/2$ \\
\hspace{24mm} \textbf{if} $\Delta \mathsf{s}_{n} \leq \Delta \mathsf{s}_{min}$: Set $\tilde{\mathsf{s}}^*_{k}=\mathsf{s}_n$, $\bar{p}_k^B(\tilde{\mathsf{s}}^*_{k})=p^B_{H_{k}}(s_n)$,\newline
\hspace*{28mm} $\bar{s}_k^B(\tilde{\mathsf{s}}^*_{k})=s^B_{H_{k}}(\mathsf{s}_n)$ and $\bar{a}_k^B(\tilde{\mathsf{s}}^*_{k})=a^B_{H_{k}}(\mathsf{s}_n)$, set $k=k+1$, go to \textbf{1} \\
\hspace{24mm} \textbf{else}: Keep $n=n$, go to \textbf{2a.i} \\[2pt]
\hspace{8mm} \textbf{else} (outside DtP zone): Set $\Delta \mathsf{s}_{n}=\Delta s_{n}/2$ \\
\hspace{16mm} \textbf{if} $\Delta \mathsf{\mathsf{s}}_{n} \leq \Delta \mathsf{\mathsf{s}}_{min}$: Set $\tilde{\mathsf{s}}^*_{k}=\mathsf{\mathsf{s}}_n$, $\bar{p}_k^B(\tilde{\mathsf{s}}^*_{k})=p^B_{H_{k}}(\mathsf{s}_n)$,\newline
\hspace*{20mm} $\bar{s}_k^B(\tilde{\mathsf{s}}^*_{k})=s^B_{H_{k}}(\mathsf{s}_n)$ and $\bar{a}_k^B(\tilde{\mathsf{s}}^*_{k})=a^B_{H_{k}}(\mathsf{s}_n)$, set $k=k+1$, go to \textbf{1} \\
\hspace{16mm} \textbf{else}: Keep $n=n$, go to \textbf{2a.i} \\
\bottomrule
\end{tabular}
\caption{Algorithm: Gradient Flow \& Newton-Raphson Iterations}
\label{tab:algorithm}
\end{table}

\subsection{Results}
\label{sec:results_sec}
\begin{figure}[H]
    \centering
    \begin{subfigure}{0.45\textwidth}
        \centering
        \begin{overpic}[width=\textwidth]{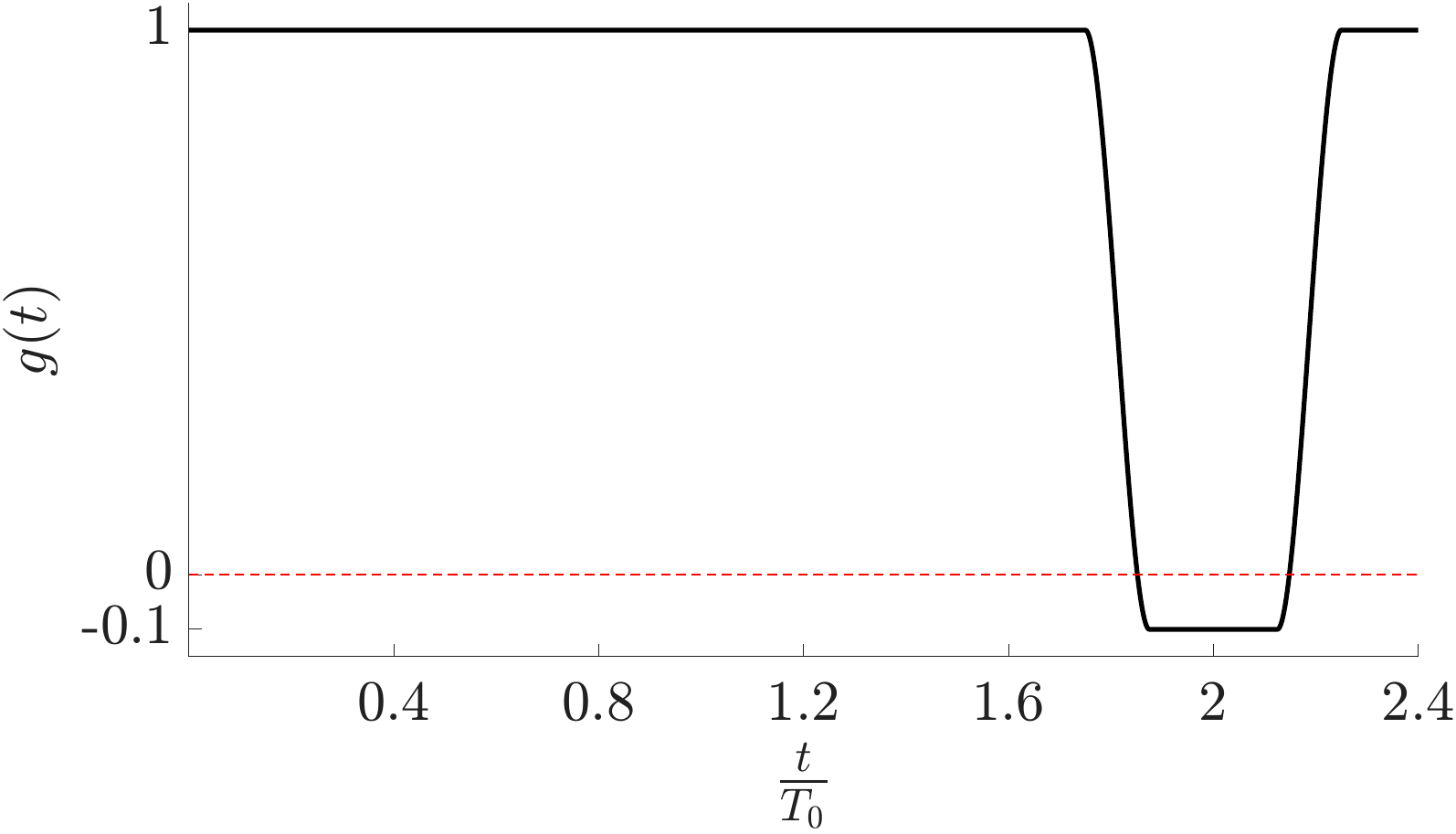}
        \put(11, 35){  % (x%, y%) desde esquina inferior izquierda
        \scalebox{0.64}{
            \begin{minipage}{0.5\textwidth}
            \fontsize{2}{1}\selectfont   % {tamaño fuente}{interlineado}
            \begin{equation*}
            g(t/T_0) = \begin{cases}
                1 & t/T_0 \le 1.75 \\[4pt]
                1-1.1\, S_{step}\!\left(\dfrac{t/T_0-1.75}{0.125}\right) & 1.75 < t/T_0 \le 1.875 \\[8pt]
                -0.1 & 1.875 < t/T_0 \le 2.125 \\[4pt]
                -0.1 + 1.1\, S_{step}\!\left(\dfrac{t/T_0-2.125}{0.125}\right) & 2.125 < t/T_0 \le 2.25 \\[4pt]
                1 & t/T_0> 2.25
            \end{cases}
            \end{equation*}
            \begin{equation*}
                S_{step}(t)=3t^2-2t^3
            \end{equation*}
            \end{minipage}
            }
        }
    \end{overpic}
    \caption{Curve $g(t)$.}
    \label{fig:c_curve_plot_2}
    \end{subfigure}
    \hfill
    \begin{subfigure}{0.45\textwidth}
        \centering
        \includegraphics[width=\textwidth]{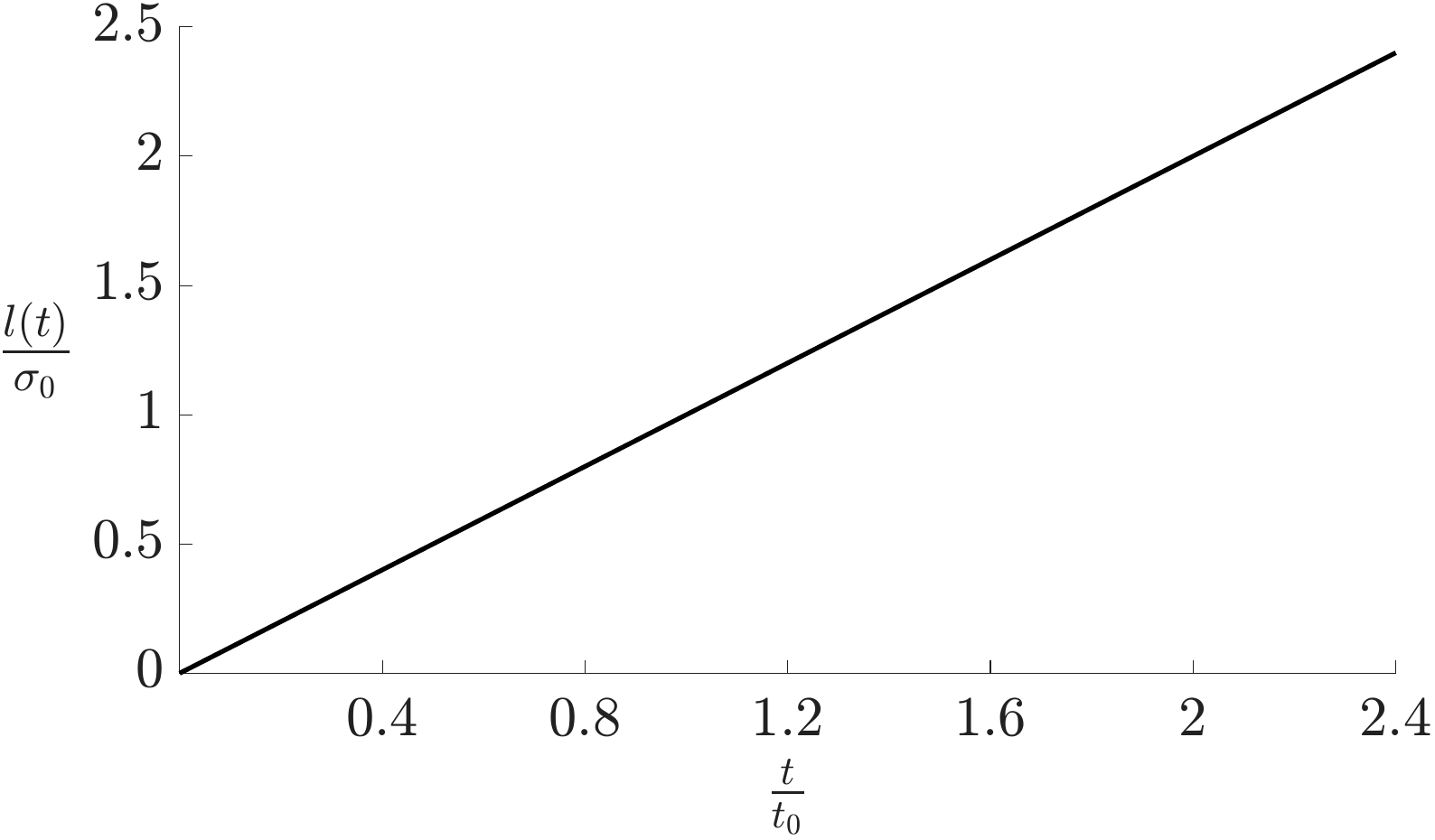}
        \caption{External load.}
        \label{fig:external_load}
    \end{subfigure}
    \caption{Curve $g(t)$ and external load applied.}
    \label{fig:external_load_and_c_curve}
\end{figure}

This Section presents numerical results for two examples, both involving the monotonically increasing external load
\begin{equation}\label{eq:c(t)}
    l(t)=l_0 \,t,
\end{equation}
where $l_0$ corresponds to the loading rate applied to the bar. The prescribed part of the plastic constitutive evolution is characterized by the response function 
\begin{equation}\label{eq:fc_generic}
    f^{c}(t)=\left(\frac{\sigma(t)}{\sigma_0}\right)^{\frac{1}{m}}c(t),
\end{equation}
where $\sigma_0$ is the initial yield stress of the material. We consider the cases $m=1$ corresponding to a rate-sensitive material, and $m=0.1$ to a rate insensitive material.
In \eqref{eq:fc_generic}, $c$ is a modulating function defined as
\begin{equation}\label{eq:c(t)dim}
    c(t) = \hat{\gamma} \, g(t),
\end{equation}
with $\hat{\gamma}$ being a dimensional constant (cf., Sec.~\ref{sec:sec2}) and $t \mapsto g(t)$ taking non-dimensional values. Here, the function $g$, shown in Fig.~\ref{fig:c_curve_plot_2}, is intentionally chosen to be strictly negative over a part of the time domain, thereby specifying a plastic behavior that would violate the Second Law \eqref{eq:2_law_red} if the evolution of plastic strain were given by $p_t = f^{c}$ (noting $l(t)$ is positive in the same domain and \eqref{eq:eq_solved}). Of course, prescribing part of the plastic strain evolution as an explicit function of time is somewhat physically unrealistic. We adopt this simplification here in order to deliberately trigger a violation on the Second Law, so as to test whether our theoretical formalism and its computational implementation are able to correct such a violation. 

Given the response function of the form \eqref{eq:fc_generic} and noting \eqref{eq:eq_solved}, $f^{c}$ becomes a specified function of time. In a more general problem, $f^{c}$ would instead be determined implicitly by the solution and cannot be directly prescribed.
Using this constitutive set up, numerical solutions for the functions $p$, $s$ and $a$ were obtained, and the total strain $u_x$ was computed from \eqref{eq:const_eqn} as $u_x = \frac{l_0 t}{E} + p$, where $E$ is the Young's modulus of the bar material.

We choose a time scale $T_0$ such that $T_0 l_0/\sigma_0 = 1$, which suffices to show the main events of interest in the simulations. The numerical solutions presented in this Section were computed with the parameter values listed in Table \ref{tab:NDparams}.
\begin{table}[H]
\centering
\caption{Parameter combinations used in the simulations.}
\label{tab:NDparams}
\renewcommand{\arraystretch}{1.2}
\begin{tabular}{cc}
\toprule
Ratio & Value \\
\midrule
$\frac{E}{\sigma_0}$  & $1\times10^3$ \\
$\frac{\hat{\gamma}}{1/T_0}$& $1\times10^{-3}$\\
$\frac{c_p}{\sigma_0}$& $1\times10^{3}$\\
$\frac{c_s}{T_0}$& $1\times10^{3}$\\
$\frac{c_a}{(\sigma_0T_0^2)}$& $1\times10^{15}$\\
\bottomrule
\end{tabular}
\renewcommand{\arraystretch}{1}
\end{table}

For each case, the approximate solutions obtained with the scheme, say $t \mapsto v(t)$, is compared with the analytical solution, say $t \mapsto v^r(t)$, computed using \eqref{eq:a(s)}, \eqref{eq:p_evol} and \eqref{eq:s2_analytic}. The percent error $v_e(t)$ \eqref{eq:error_formula} between $v$ and $v^r$ is also shown, where $m(v^r(t))$ is the mean of $v^r(t)$ in the physical time domain and $|v^r(t)|$ is the absolute value of $v^r$ at the time instant $t$.
\begin{equation}\label{eq:error_formula}
    v_e(t) = \begin{cases}
       \frac{v(t)-v^r(t)}{m(v^r(t))}\times100 \ [\%], &\qquad |v^r(t)| \le 1\times10^{-3} \\[4pt]
       \frac{v(t)-v^r(t)}{v^r(t)}\times100 \ [\%], &\qquad |v^r(t)| > 1\times10^{-3}. \\ 
    \end{cases}
\end{equation}

An important point to note about our numerical solution procedure relying on a sequence of convex optimization problems is as follows. Strictly speaking, the numerical scheme does not attempt to discretize the problem defined by \eqref{eq:min}-\eqref{eq:a_s_p}, where the solution follows from minimizing $\mathcal{H}$ with the base state $\bar{U} = (0,0,0)$. The scheme instead works with a sequence of functionals $S_H$, parametrized by a corresponding sequence of base states defined by the algorithm. The initial base states for all simulations presented were chosen as $\bar{U}=[0,0.1,0]$. The base state $\bar{s} = 0$ is a natural choice to guide the solution to that of \eqref{eq:min}-\eqref{eq:a_s_p}. However, since the DtP map for $s$ is $s_H=c_s\bar{s}/(\alpha +c_s)$, setting $\bar{s}=0$ would force $s_H=0$ throughout the discrete algorithm, trapping it at this starting value. Consequently, a small non-zero starting $\bar{s}$ is used, while remaining close to desired guess. The effect of larger initial values of $\bar{s}$ is discussed further in this Section. Our numerical experiments, thus, also test if the scheme is able to reproduce the analytical result for the problem \eqref{eq:min}-\eqref{eq:a_s_p}, despite this difference in the formulation.

\begin{figure}[H]
    \centering
    \begin{subfigure}{0.45\textwidth}
        \centering
        \includegraphics[width=\textwidth]{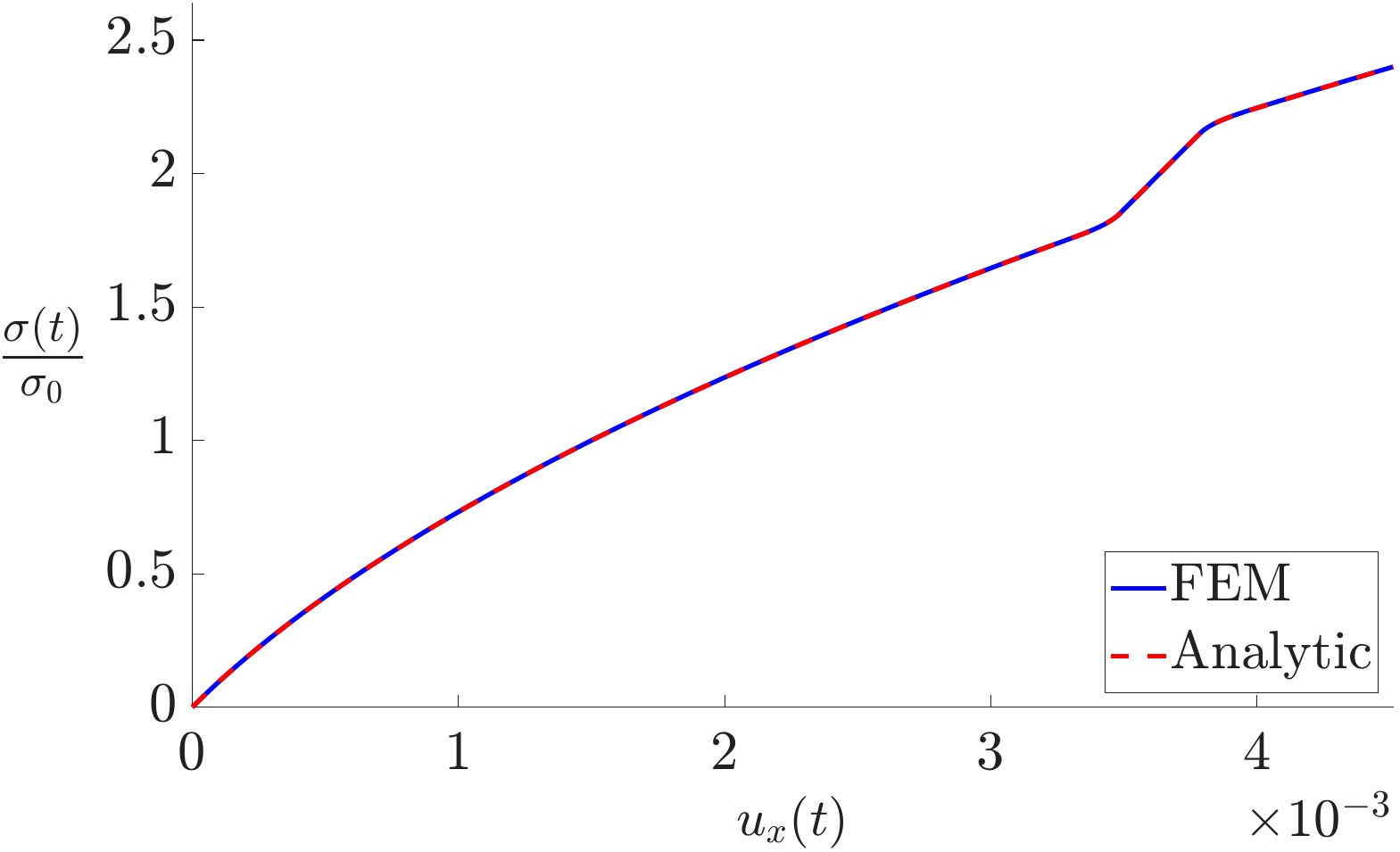}
        \caption{Stress-strain curve for $m=1$.}
        \label{fig:ss1}
    \end{subfigure}
    \hfill
    \begin{subfigure}{0.45\textwidth}
        \centering
        \includegraphics[width=\textwidth]{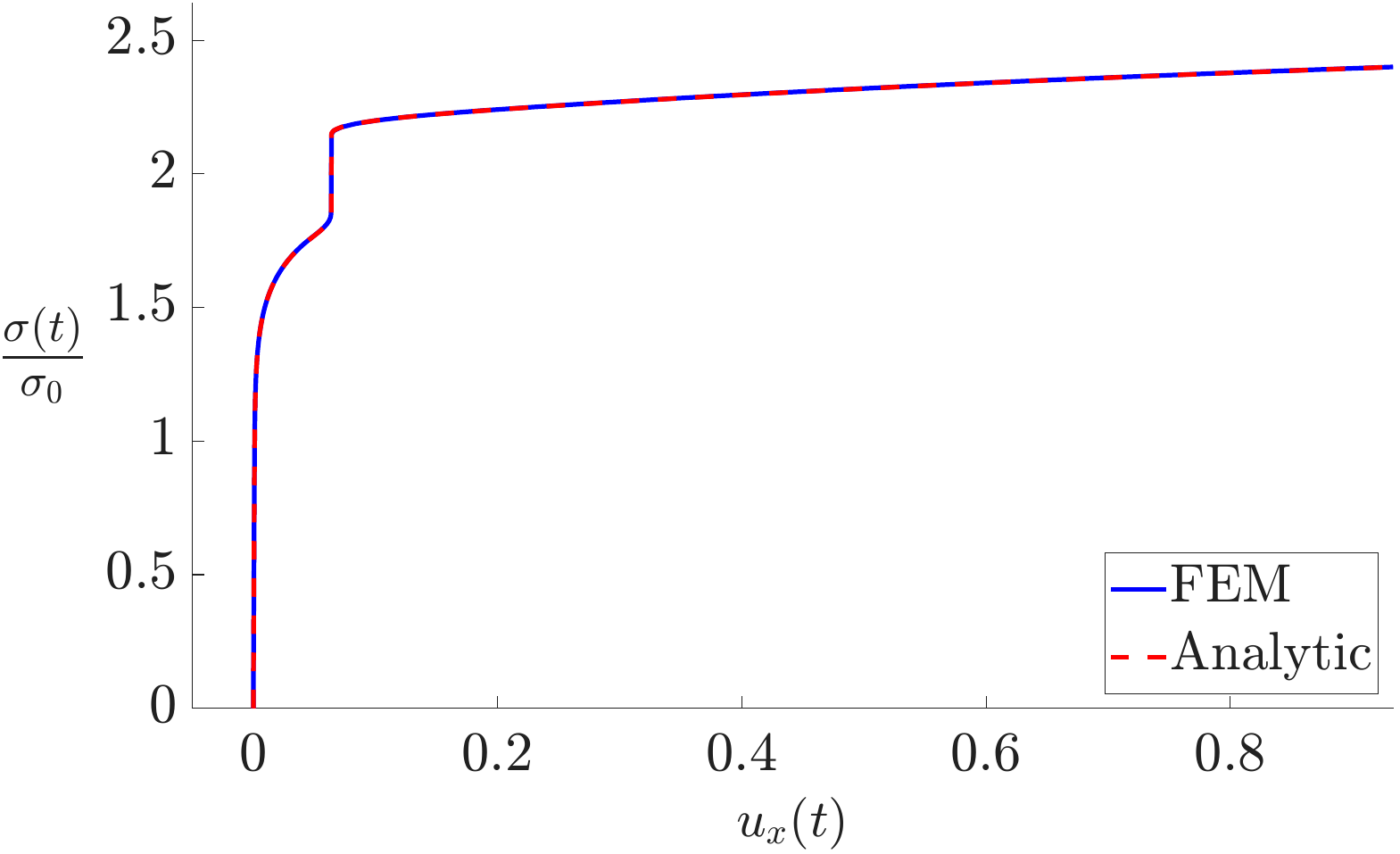}
        \caption{Stress-strain curve for $m=0.1$.}
        \label{fig:ss2}
    \end{subfigure}
    \caption{Stress-strain relations.}
    \label{fig:ss_both_cases}
\end{figure}

\begin{figure}[H]
    \centering
    \begin{subfigure}{0.45\textwidth}
        \centering
        \includegraphics[width=\textwidth]{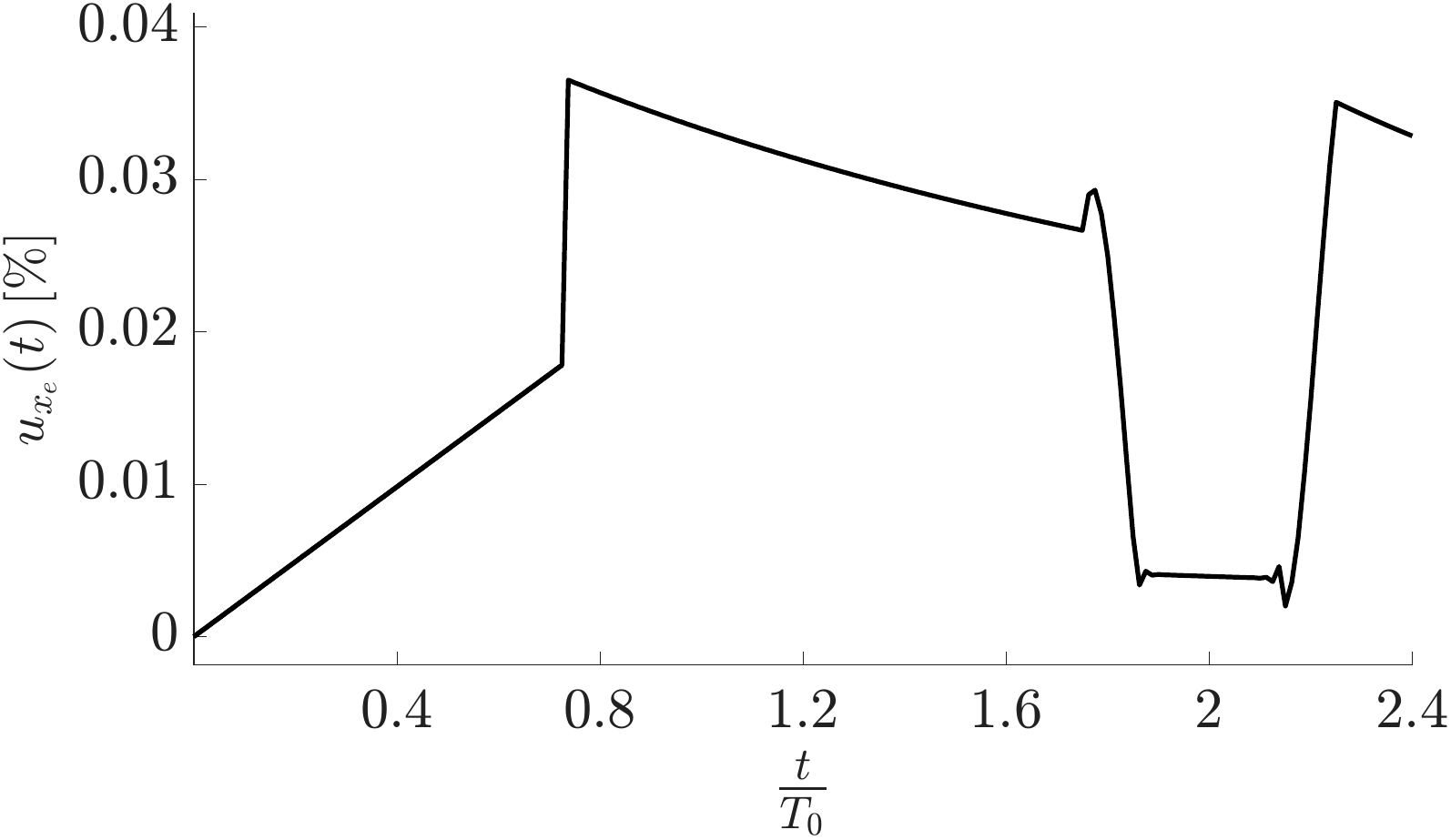}
        \caption{Error for $m=1$.}
        \label{fig:error_ss1}
    \end{subfigure}
    \hfill
    \begin{subfigure}{0.45\textwidth}
        \centering
        \includegraphics[width=\textwidth]{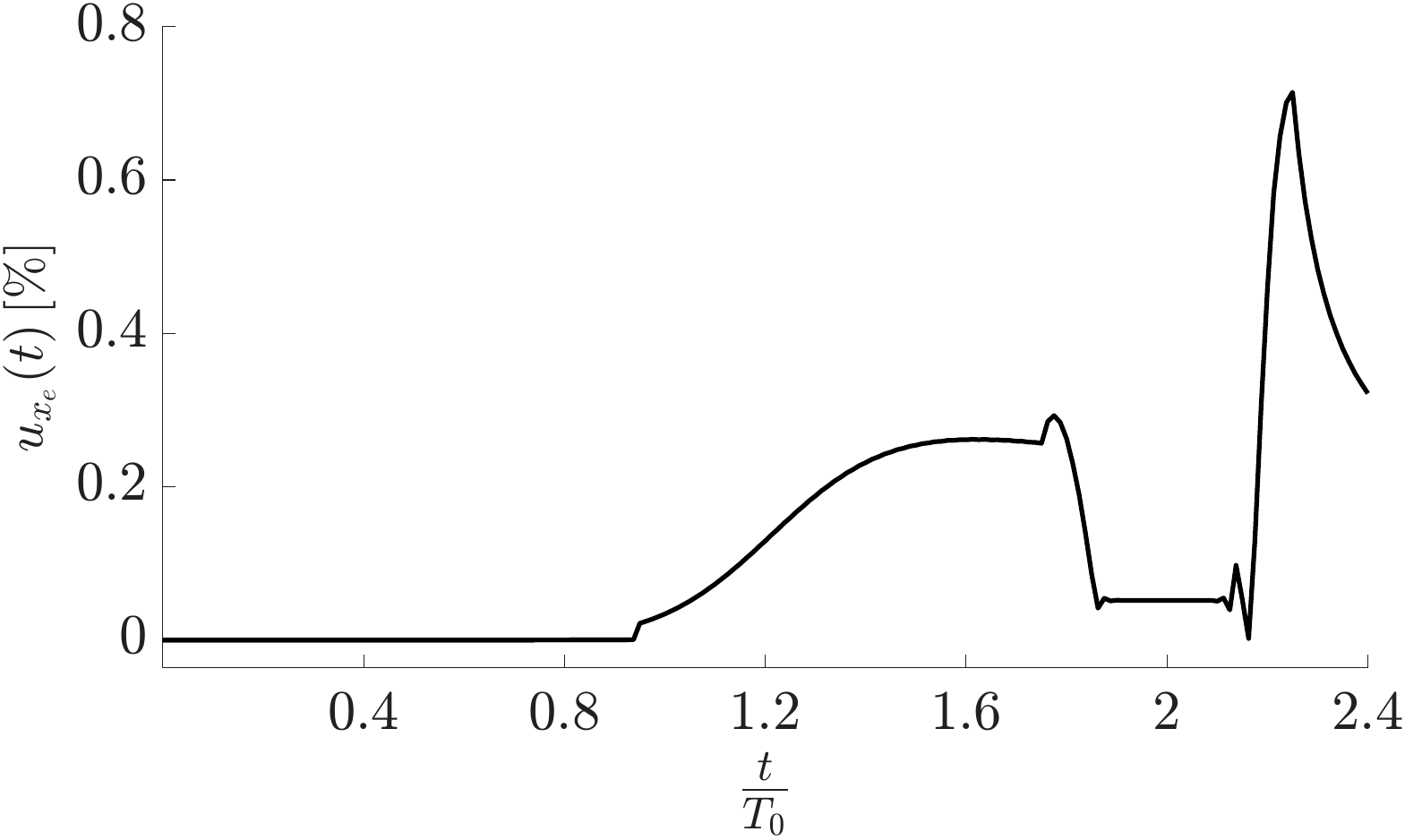}
        \caption{Error for $m=0.1$.}
        \label{fig:error_ss2}
    \end{subfigure}
    \caption{Error percentage for total strain.}
    \label{fig:error_ss}
\end{figure}

The stress-strain curves obtained for the cases $m=1$ case and $m=0.1$ are presented in Figs. \ref{fig:ss1} and \ref{fig:ss2}, with the corresponding error for $u_x$ shown in Figs. \ref{fig:error_ss1} and \ref{fig:error_ss2}. Both curves exhibit an apparent `elastic gap,' where, despite the external load increasing monotonically, the stiffness reverts to a purely elastic value $E$. This behavior is not prescribed, but emerges naturally from the scheme through the activation of the corrective variable $a$ which enforces the Second Law \eqref{eq:2_law_red}, as discussed next.

\begin{figure}[H]
    \centering
    \begin{subfigure}{0.45\textwidth}
        \centering
        \includegraphics[width=\textwidth]{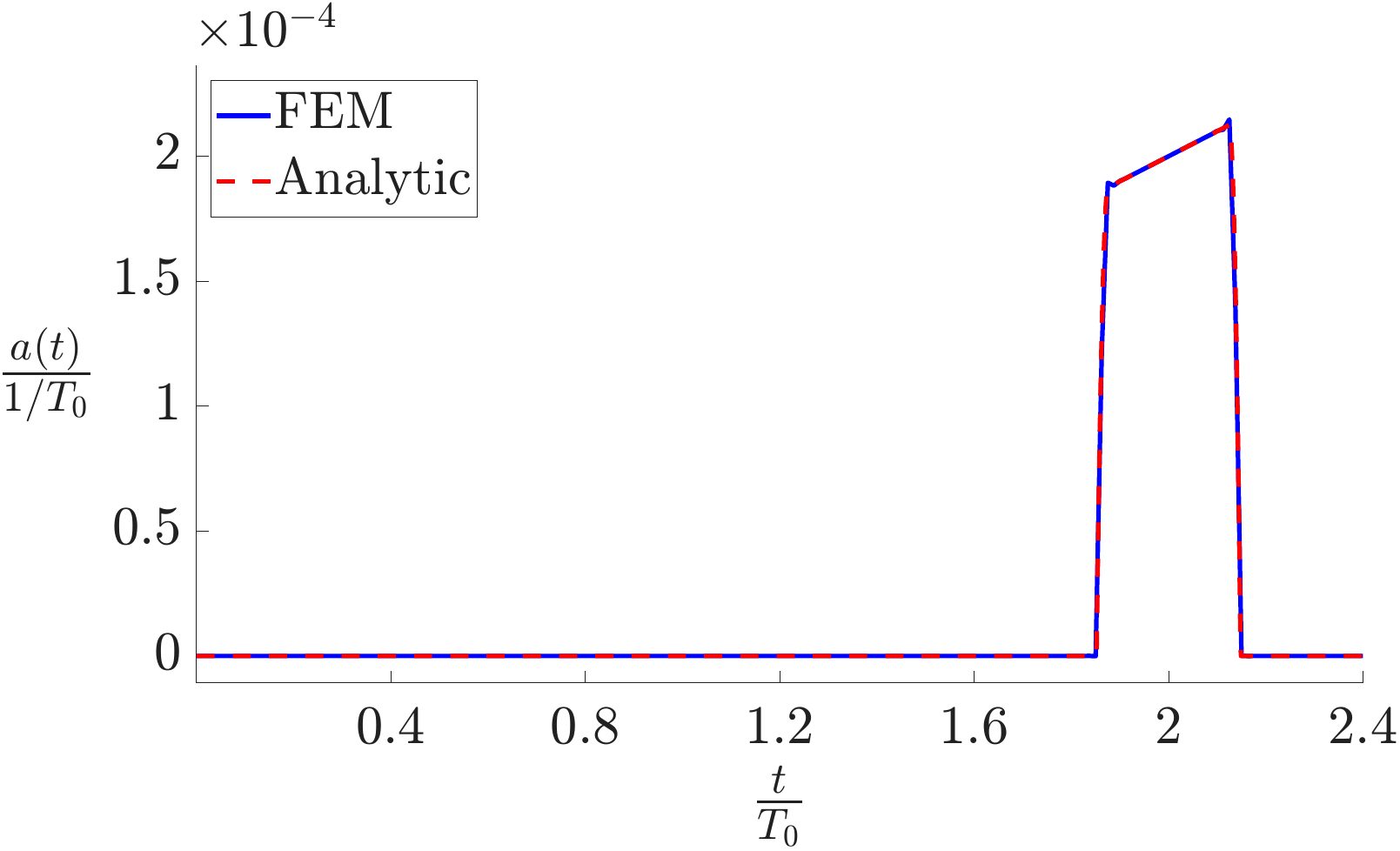}
        \caption{$a(t)$ for $m=1$.}
        \label{fig:ah1}
    \end{subfigure}
    \hfill
    \begin{subfigure}{0.45\textwidth}
        \centering
        \includegraphics[width=\textwidth]{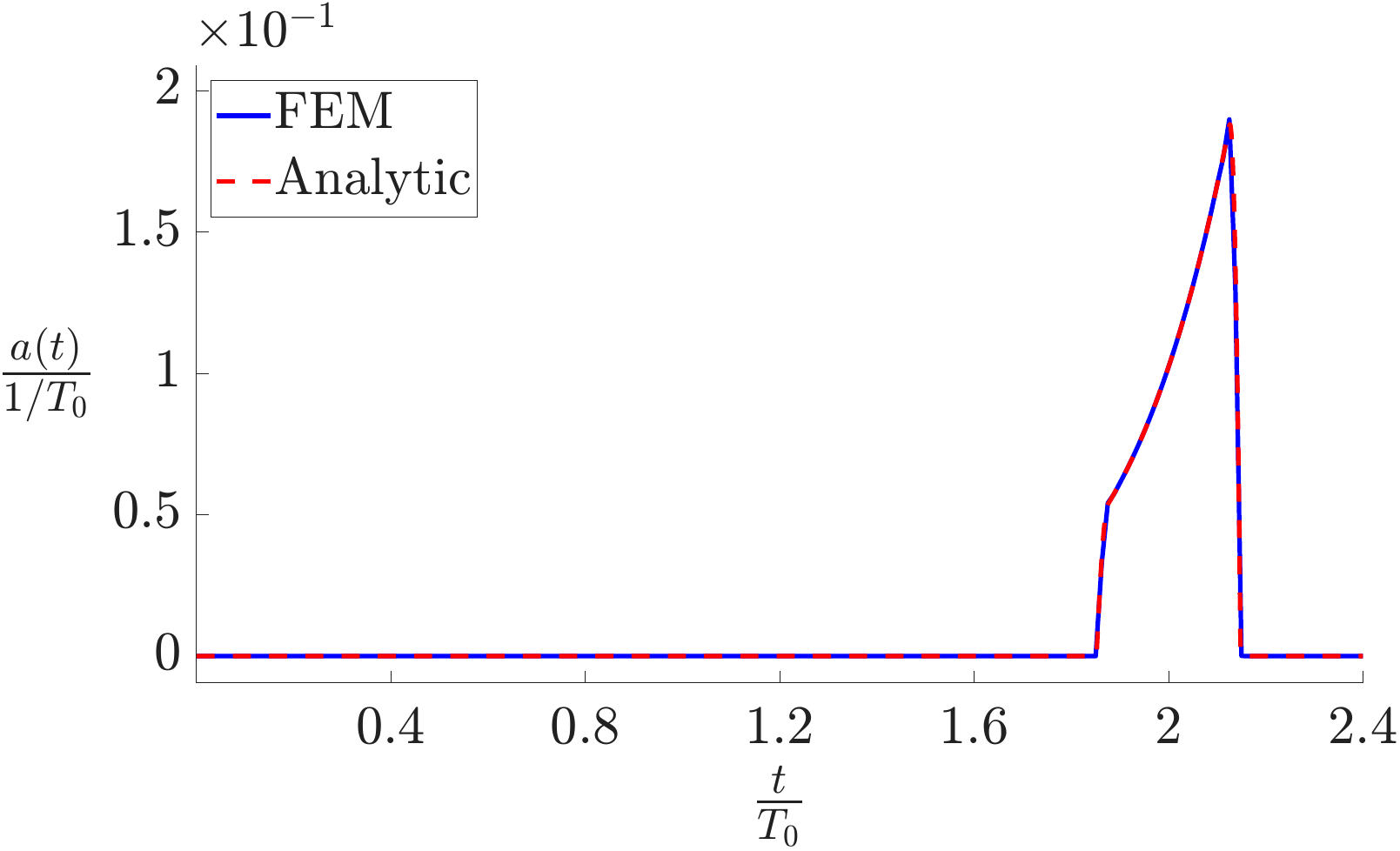}
        \caption{$a(t)$ for $m=0.1$.}
        \label{fig:ah2}
    \end{subfigure}
    \caption{Function $a(t)$.}
    \label{fig:ah_solution1}
\end{figure}

\begin{figure}[H]
    \centering
    \begin{subfigure}{0.45\textwidth}
        \centering
        \includegraphics[width=\textwidth]{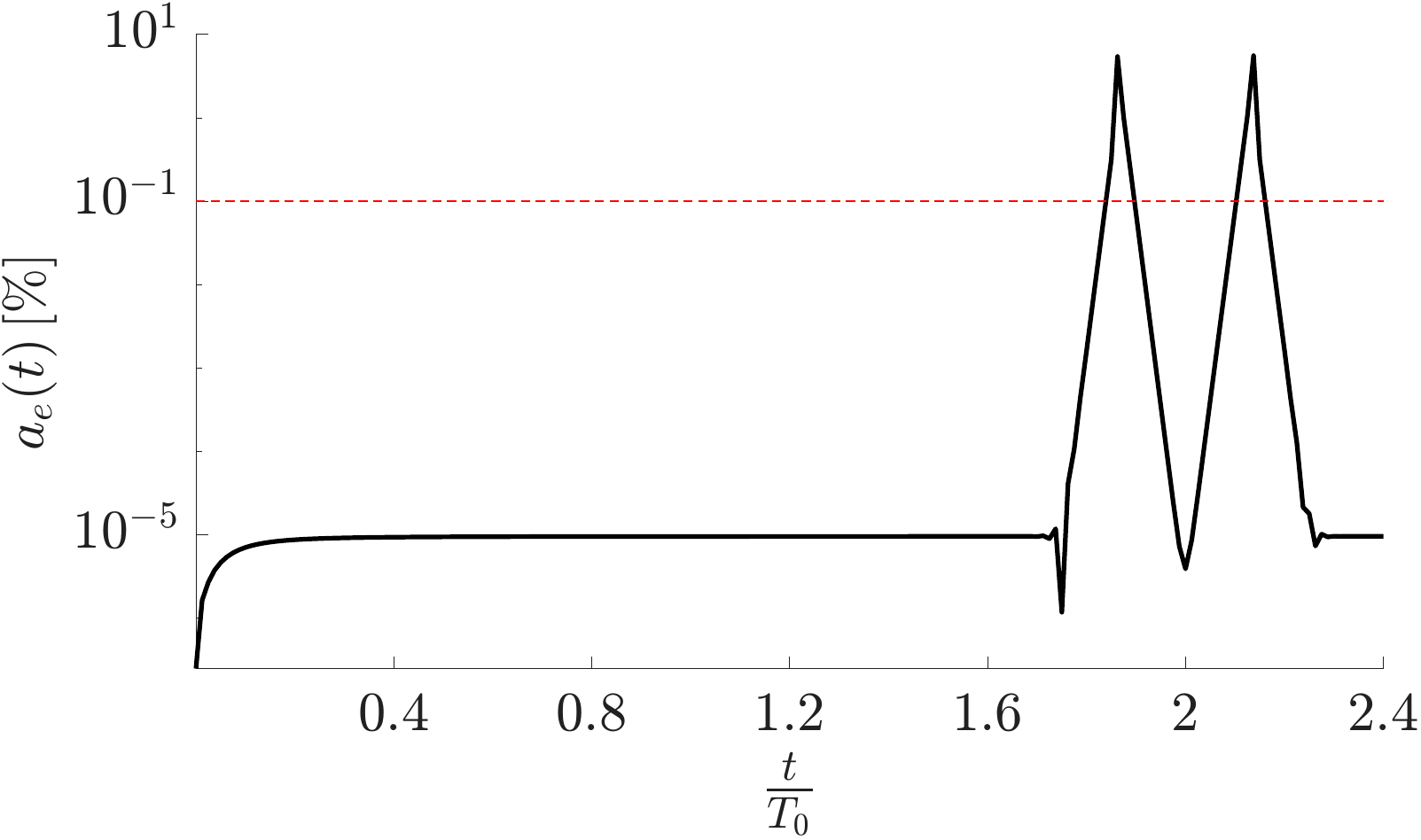}
        \caption{Error for $m=1$.}
        \label{fig:error_ah1}
    \end{subfigure}
    \hfill
    \begin{subfigure}{0.45\textwidth}
        \centering
        \includegraphics[width=\textwidth]{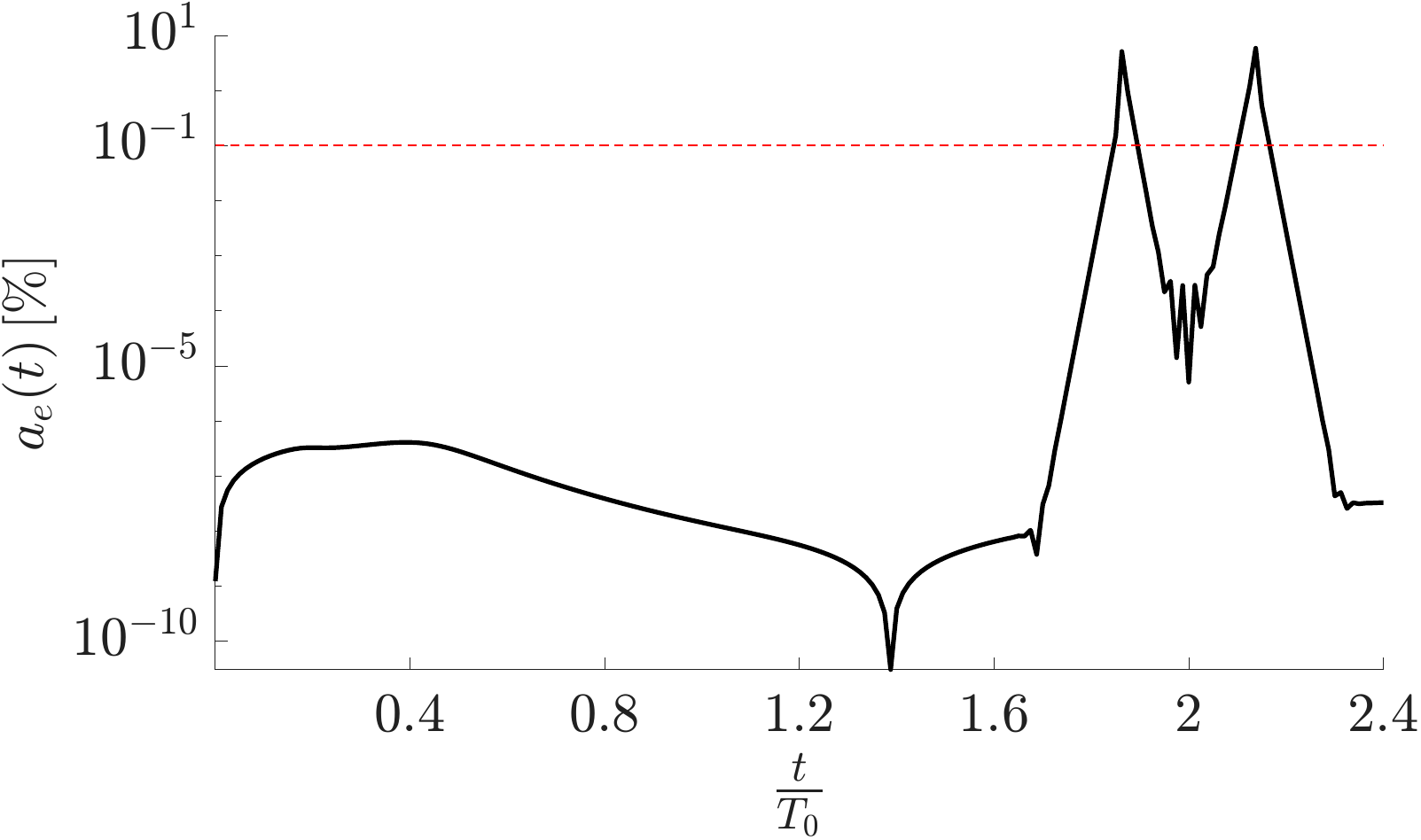}
        \caption{Error for $m=0.1$.}
        \label{fig:error_ah2}
    \end{subfigure}
    \caption{Error percentage for $a(t)$.}
    \label{fig:error_ah}
\end{figure}

\begin{figure}[H]
    \centering
    \begin{subfigure}{0.45\textwidth}
        \centering
        \includegraphics[width=\textwidth]{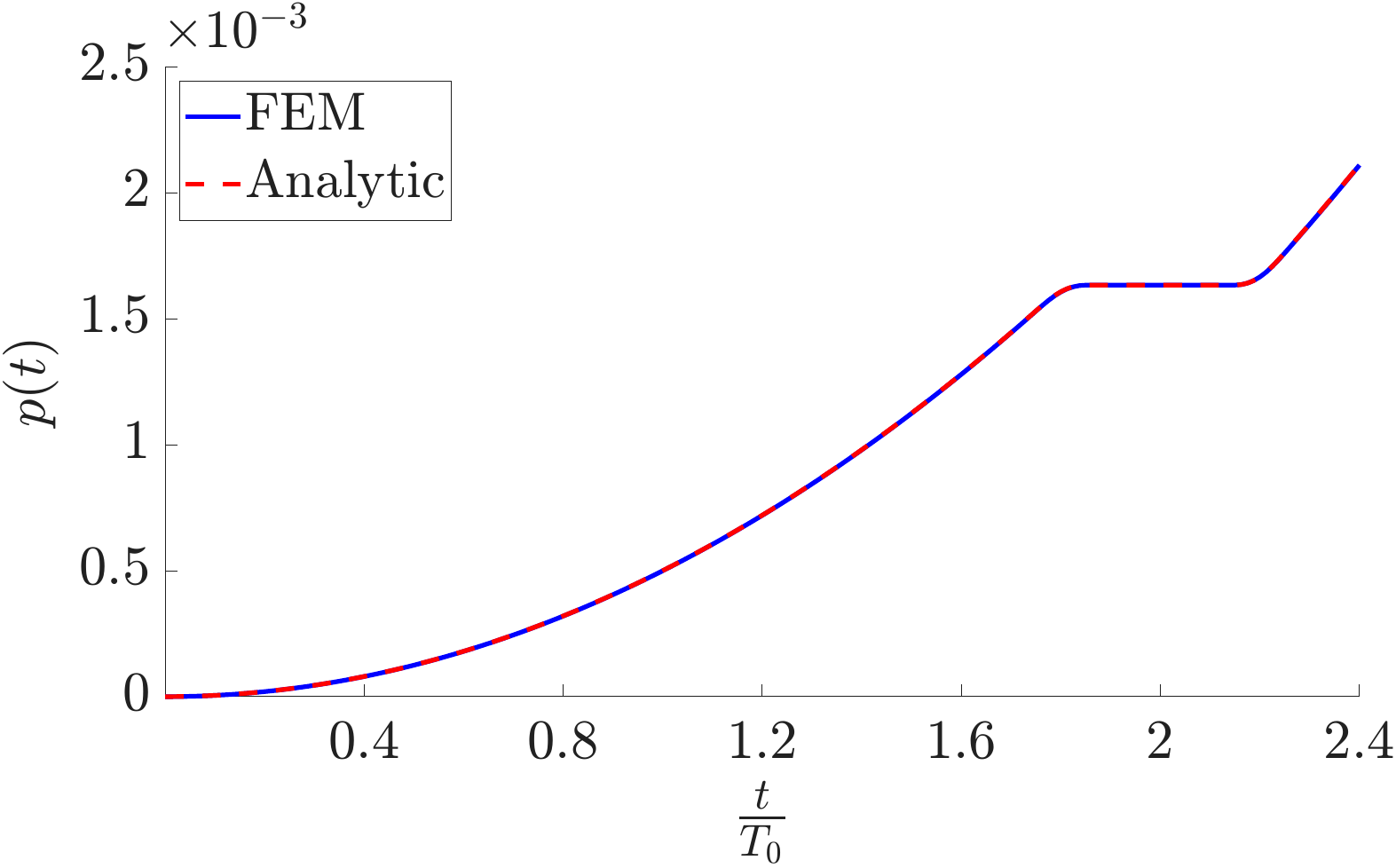}
        \caption{Plastic strain for $m=1$.}
        \label{fig:ph1}
    \end{subfigure}
    \hfill
    \begin{subfigure}{0.45\textwidth}
        \centering
        \includegraphics[width=\textwidth]{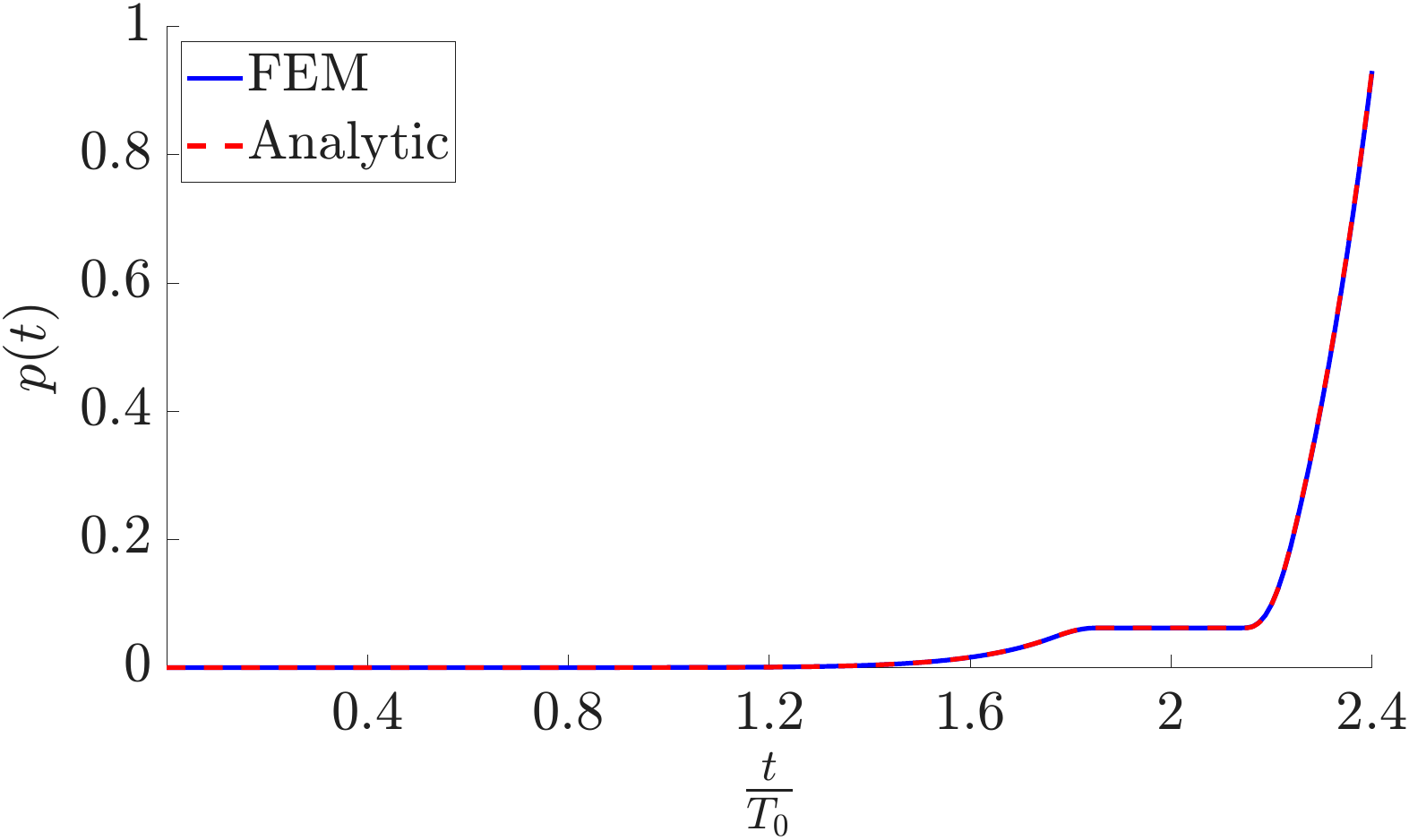}
        \caption{Plastic strain for $m=0.1$.}
        \label{fig:ph2}
    \end{subfigure}
    \caption{Plastic strain.}
    \label{fig:ph_solution1}
\end{figure}

\begin{figure}[H]
    \centering
    \begin{subfigure}{0.45\textwidth}
        \centering
        \includegraphics[width=\textwidth]{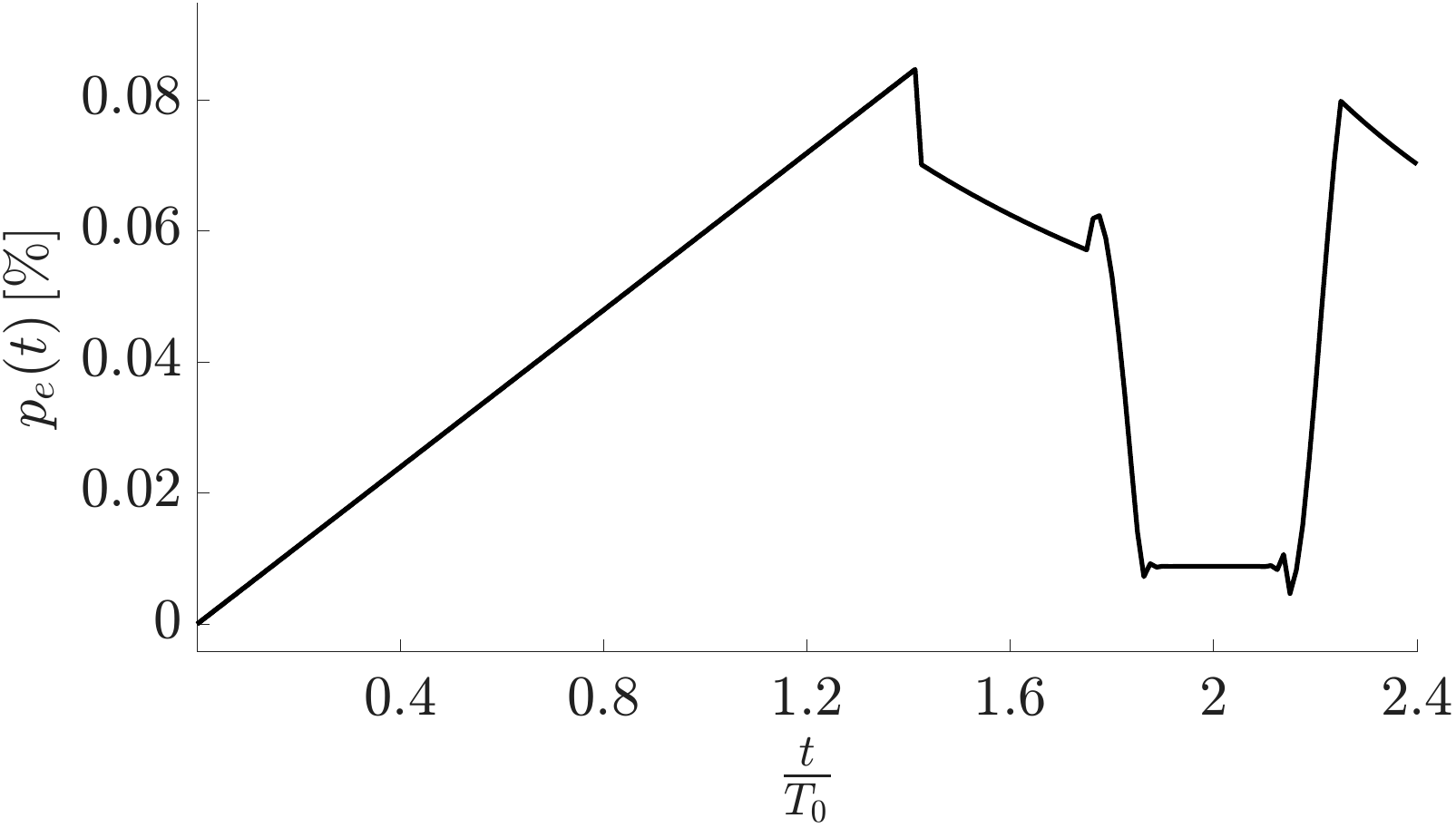}
        \caption{Error for $m=1$.}
        \label{fig:error_ph1}
    \end{subfigure}
    \hfill
    \begin{subfigure}{0.45\textwidth}
        \centering
        \includegraphics[width=\textwidth]{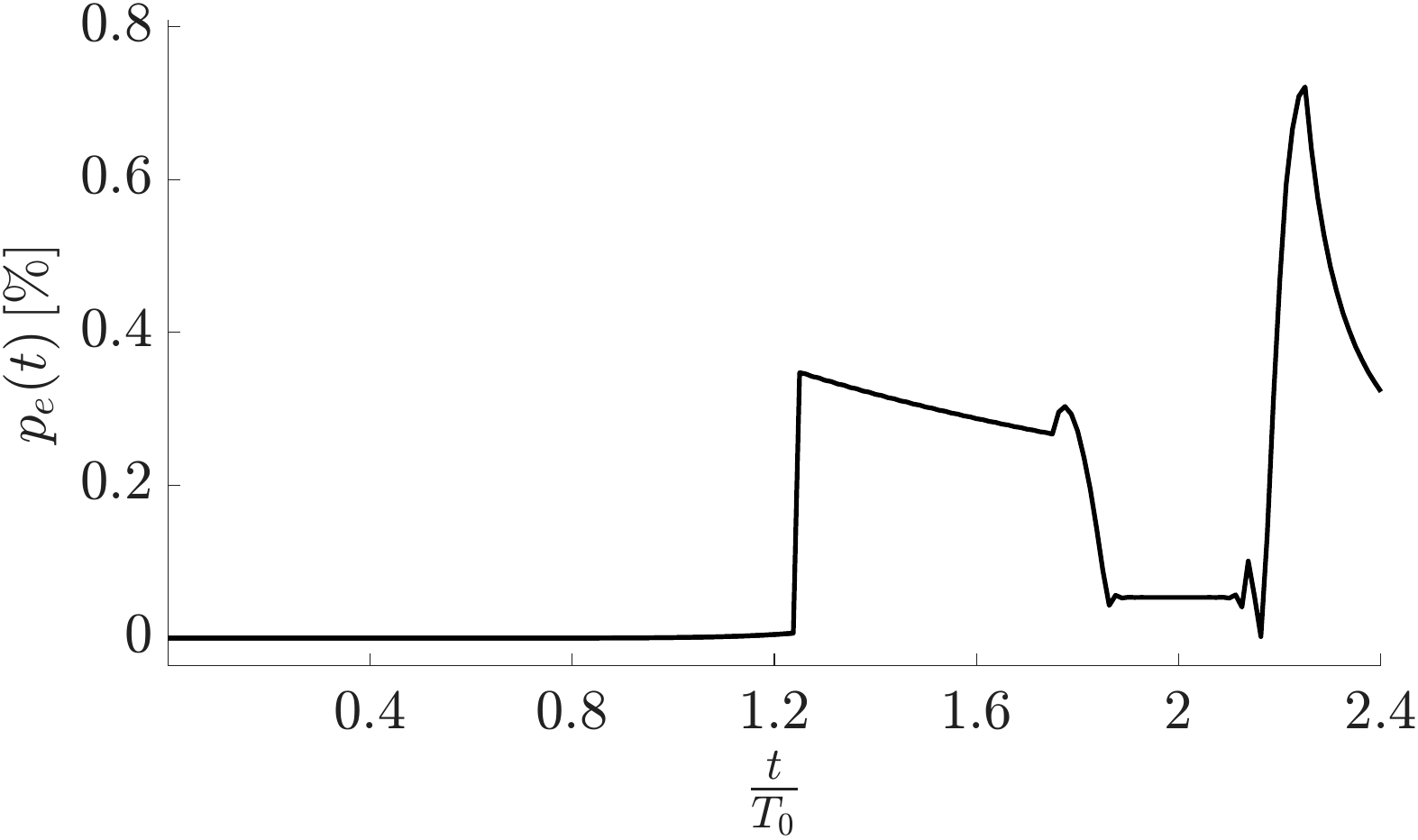}
        \caption{Error for $m=0.1$.}
        \label{fig:error_ph2}
    \end{subfigure}
    \caption{Error percentage for the plastic strain.}
    \label{fig:error_ph}
\end{figure}

\begin{figure}[H]
    \centering
    \begin{subfigure}{0.45\textwidth}
        \centering
        \includegraphics[width=\textwidth]{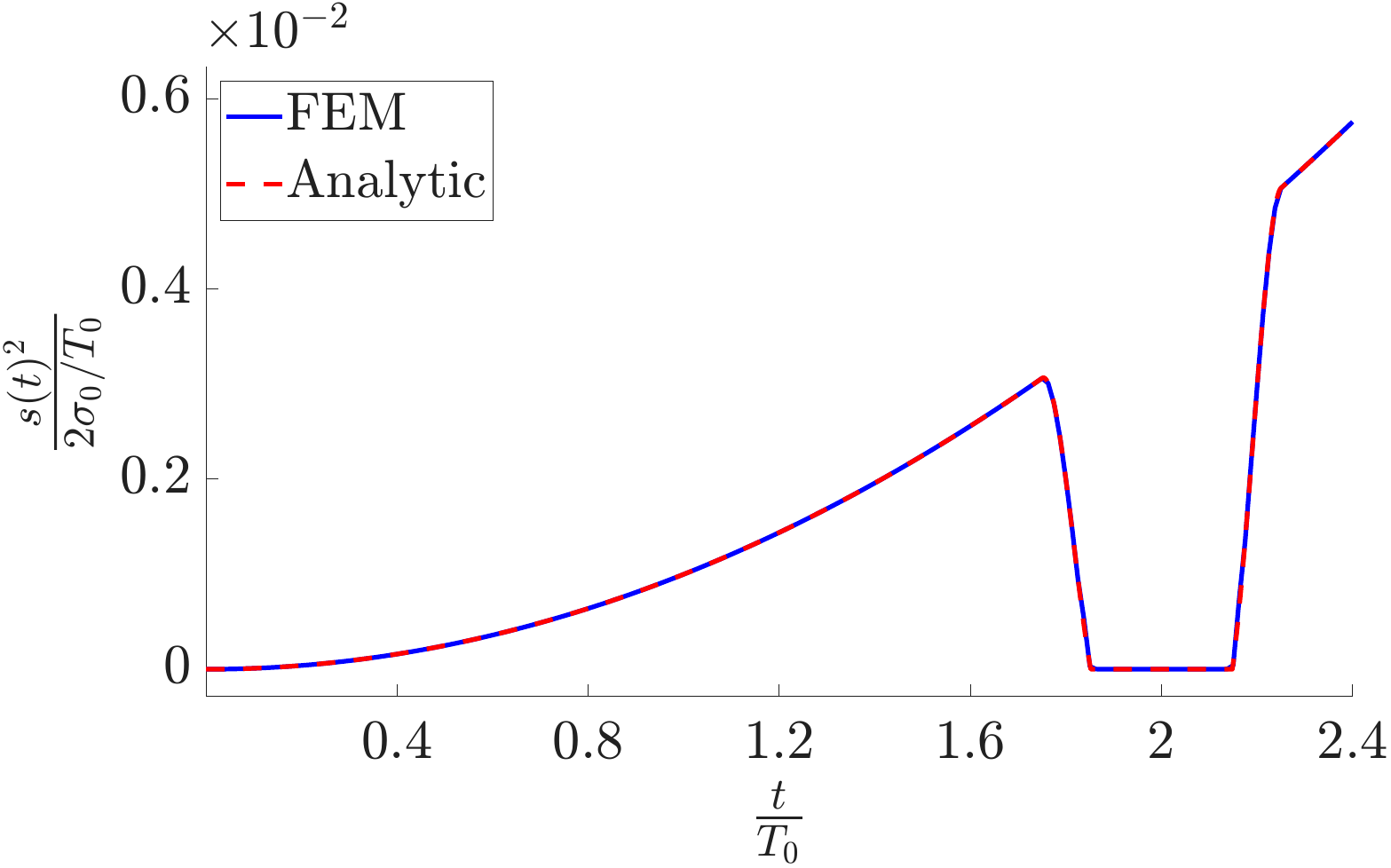}
        \caption{Dissipated energy for $m=1$.}
        \label{fig:sh1_energy}
    \end{subfigure}
    \hfill
    \begin{subfigure}{0.45\textwidth}
        \centering
        \includegraphics[width=\textwidth]{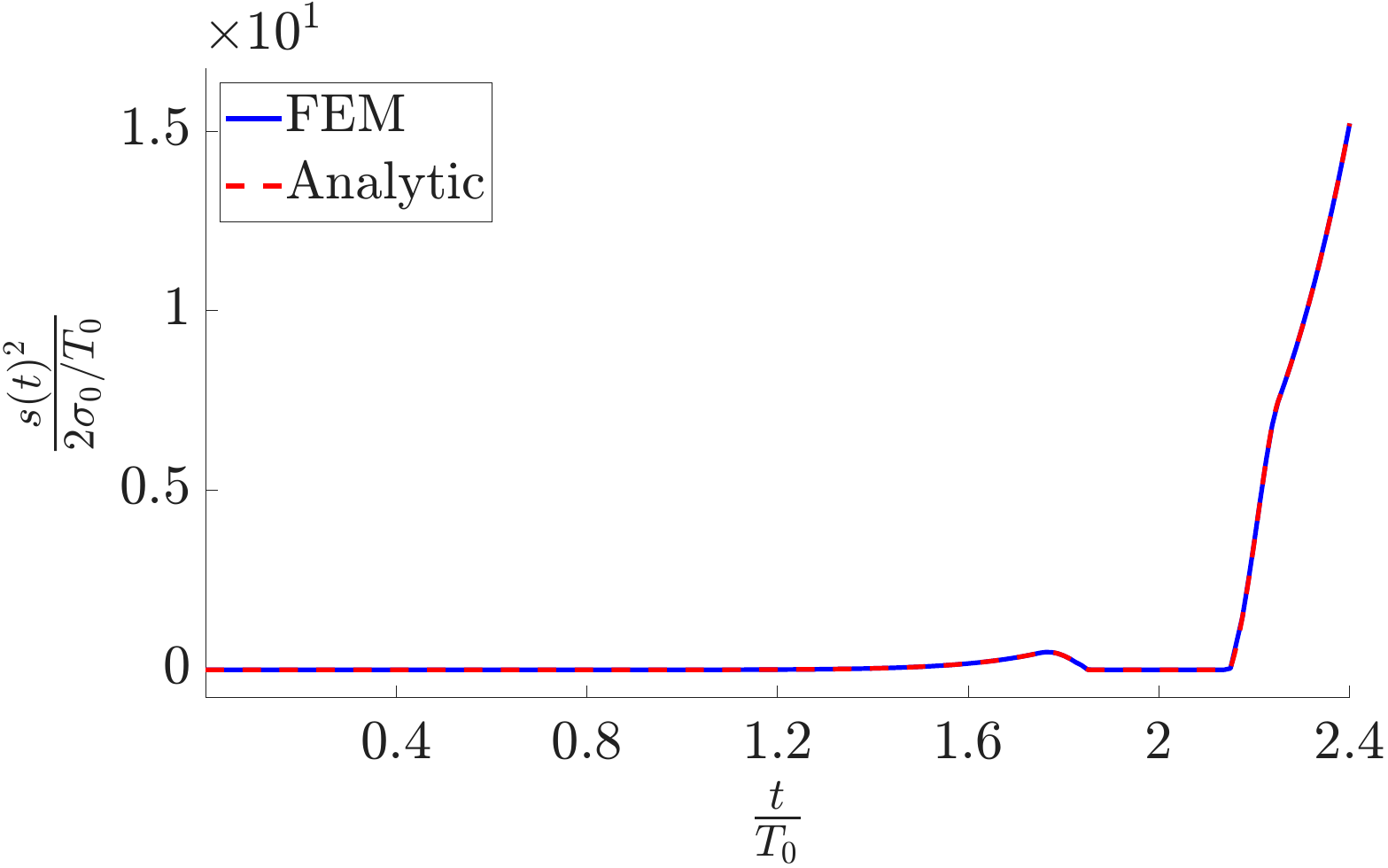}
        \caption{Dissipated energy for $m=1$ case.}
        \label{fig:sh2_energy}
    \end{subfigure}
    \caption{Dissipated energy.}
    \label{fig:sh_energy}
\end{figure}

\begin{figure}[H]
    \centering
    \begin{subfigure}{0.45\textwidth}
        \centering
        \includegraphics[width=\textwidth]{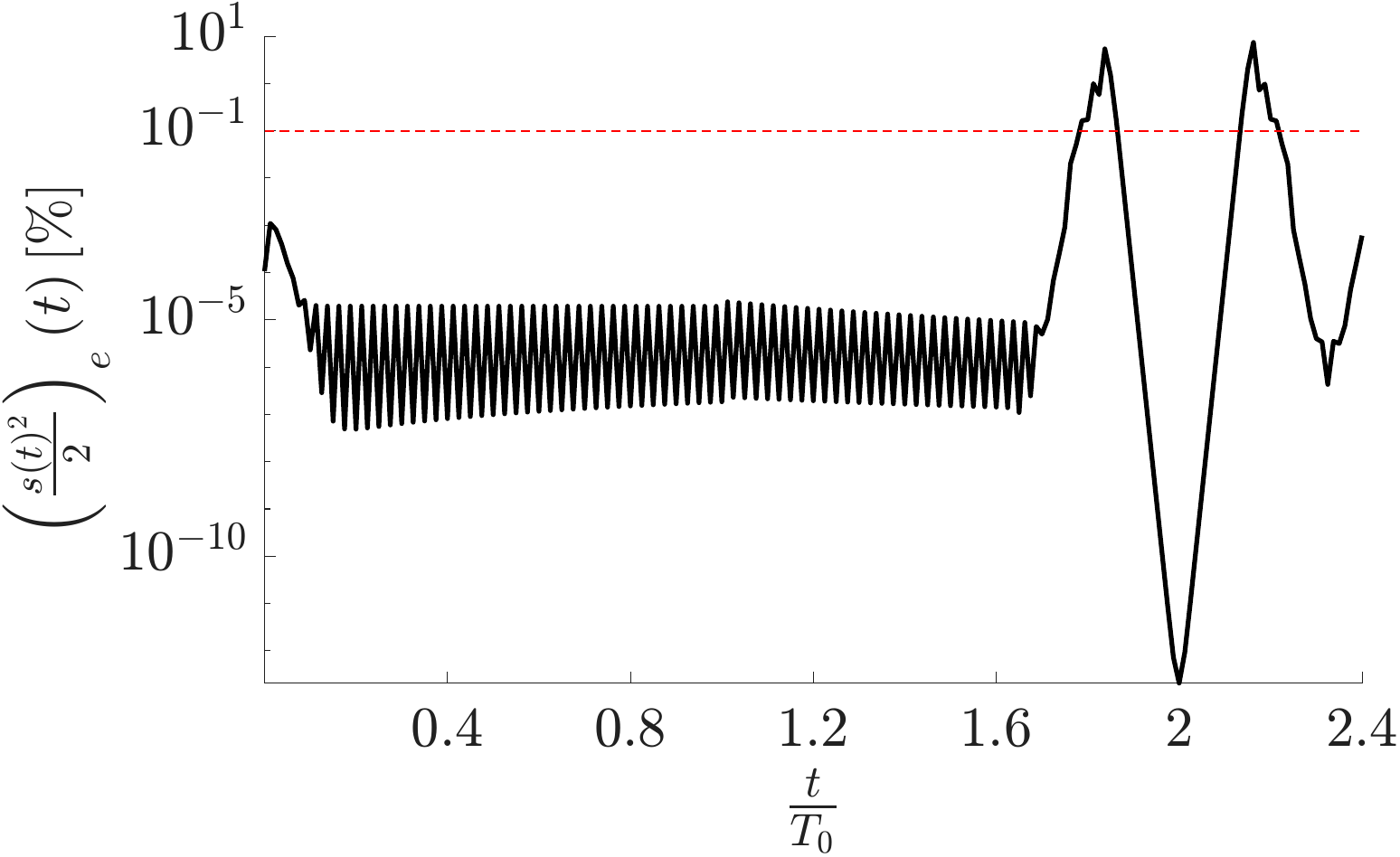}
        \caption{Error for $m=1$.}
        \label{fig:error_energy1}
    \end{subfigure}
    \hfill
    \begin{subfigure}{0.45\textwidth}
        \centering
        \includegraphics[width=\textwidth]{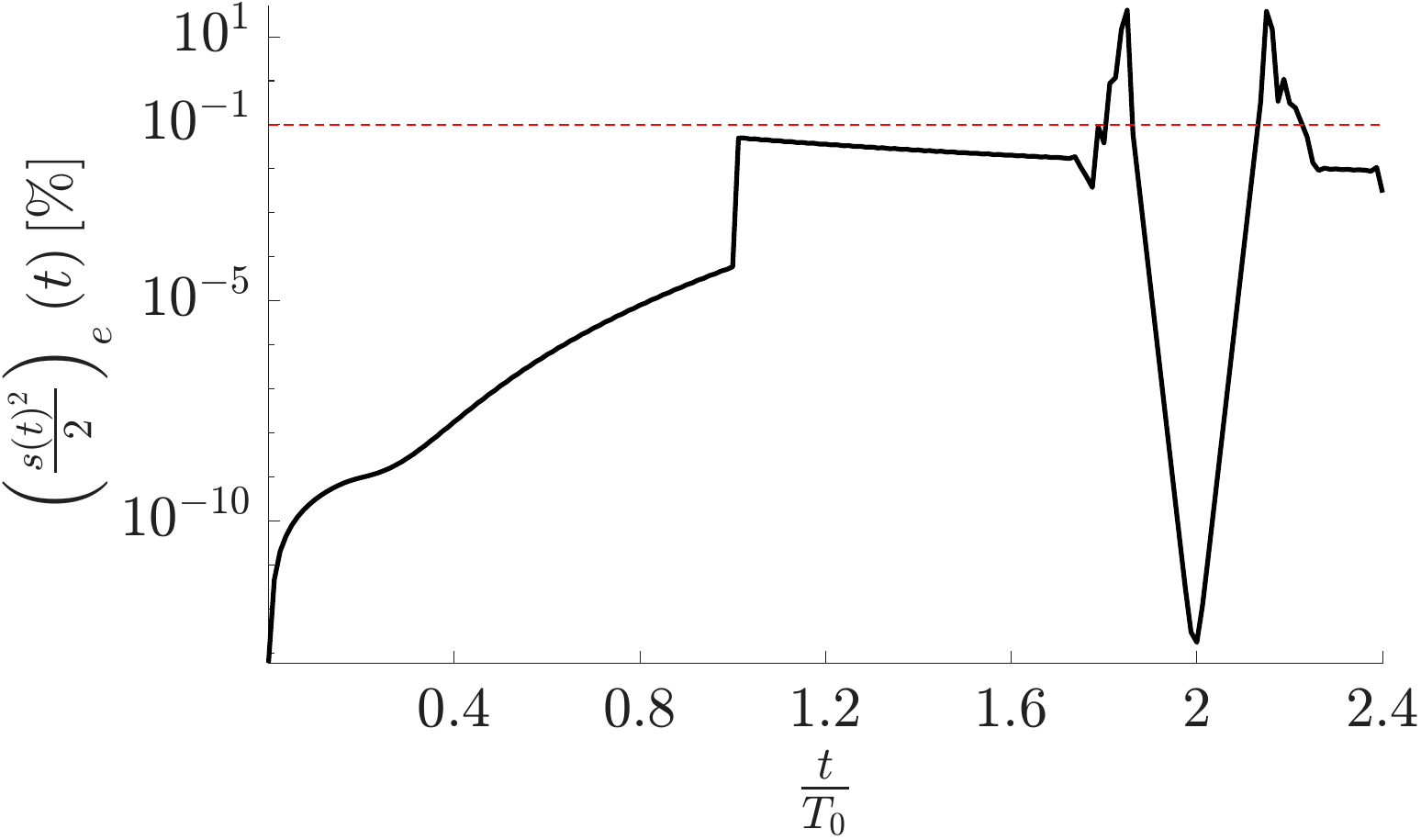}
        \caption{Error for $m=0.1$.}
        \label{fig:error_energy2}
    \end{subfigure}
    \caption{Error percentage for the dissipated energy.}
    \label{fig:error_energy}
\end{figure}

The solution for $a(t)$ is shown in Figs. \ref{fig:ah1} to \ref{fig:error_ah2}. Recall that the constitutive response for the plastic strain rate in this problem is composed of two parts, a completely determined part $t \mapsto f^c(t)$ from the boundary condition and  \eqref{eq:fc_generic} defined through a power law and modulated with the function $g$ (shown in \ref{fig:c_curve_plot_2}), and a corrective part $a$, which is instead determined by the scheme. In the time interval where $g$ becomes negative, the prescribed part $f^c$ would drive the dissipation negative, directly violating the Second Law constraint \eqref{eq:2_law_red}, were it to operate in isolation. To correct this, the scheme activates $a(t)>0$ within that interval, modifying the constitutive response so as to satisfy the Second Law.
\noindent The effect of this correction is that the constitutive response \eqref{eq:constit_eq_fc_+_a} becomes $p_t = 0$ in the activation interval. This is shown in Figs. \ref{fig:ph1} to \ref{fig:error_ph2}, where $p$ remains constant over the interval where $a(t)>0$ and $f^c<\frac{c_s l}{c_a}$. Thus, $\frac{d\sigma}{dt} = E \frac{du_x}{dt}$ in this interval, and since $u$ is a monotonically increasing function of $t$ here, we recover the elastic slope $\frac{d \sigma}{du_x} = E$ in the regions with the elastic gaps of Figs. \ref{fig:ss1} and \ref{fig:ss2}. Note that the threshold $\frac{c_sl}{c_a}$ for $f^c$ \eqref{eq:s2_analytic} is small but positive in the present simulations since $\frac{c_s \sigma_0}{c_a \hat{\gamma}} = 1 \times 10^{-9}$ (see Table \ref{tab:NDparams}), so the activation interval for $a$, shown in Fig.~\ref{fig:ah1}, closely resembles the limit case for $f^c<0$ in \eqref{eq:s2_limit}.
Looking at the dissipated energy, $E_{diss}:=s^2/2$, shown in Figs. \ref{fig:sh1_energy} to \ref{fig:error_energy2}, it is confirmed that the dissipation remains strictly non-negative throughout the entire time domain for both cases, vanishing only when $p_t=0$ and recovering positive values once plasticity resumes.

It is also worth noting that, in the interval where the response function is deliberately modified to violate the Second Law, the scheme corrects it through $a$ by a minimal amount required to satisfy \eqref{eq:2_law_red}, i.e., to drive the dissipation to precisely zero, rather than to any positive value. This behavior is explained by the large cost that any value of $a\neq0$ incurs in the dual functional $S_H[D]$, through the potential $H$ and the corresponding weight $c_a$, which was chosen to be large in this simulations.

Regarding numerical accuracy, as shown in Fig. \ref{fig:error_ss}, the scheme achieves a percent error below $0.05\%$ for the stress-strain relation in the $m=1$ case, and below $0.8\%$ in the $m=0.1$ case. The errors for $p$, shown in Fig. \ref{fig:error_ph}, follow the same trend, staying below $0.1\%$ and $1\%$ for both respective cases. As shown in Fig. \ref{fig:error_ah}, the error for $a$ remains under $0.1\%$ over most of the domain, with two localized spikes of up to $6\%$ occurring where $g$ rapidly transitions between its plateau values (see Fig. \ref{fig:c_curve_plot_2}). A similar pattern is observed in the dissipated energy curves, where the error is below $0.1\%$ except for two spikes reaching values up to $40\%$ in the $m=0.1$ case. In both cases, the largest error is localized precisely at those rapid transitions, suggesting that they originate from mesh resolution in that region of the time domain, rather than a cumulative numerical error from the scheme. It is expected that using a finer mesh, or a smoother transition in $g$, would help reduce the error in these regions. Of course, a more adapted discretization like a Discontinuous Galerkin scheme for the dual fields would also be appropriate, at higher cost.

It is important to clarify that the numerical scheme recovers a solution that resembles the one analytically obtained in Sec.~\ref{sec:sec2} which follows from minimizing $\mathcal{H}$, with $\bar{U} = (0,0,0)$ even though the numerical scheme instead minimizes $H_k$ parametrized by a sequence of base states and, as explained above, requires a non-zero initial value for $\bar{s}$. The correspondence observed in the solutions is not a coincidence, and is in fact due to the small value used for the initial $\bar{s}$. Numerical experiments (not shown) with initial values of $\bar{s}$ of the order of $1\times10^6$ converge to a different solution of the primal system, still satisfying force equilibrium, the constitutive equation, and the Second Law, but does not resemble the analytical solution obtained by minimizing $\mathcal{H}$. This confirms that the initial value of the base state is acting as a selection parameter among infinite possible solutions, and that the small value of $\bar{s}$ used here guides the scheme toward the desired closed-form solution, within the errors reported above.

\section{Conclusion}
In this work we demonstrate the scheme presented in \cite{ach3, acharya2025second, AV_nash} through a specific example: the quasi static response of a rate-dependent elastic-plastic bar whose prescribed constitutive equation was intentionally chosen to violate the Dissipation Inequality. Closed-form and numerical solutions were developed for this problem.

The results show that the proposed scheme is capable of correcting the prescribed constitutive response, through the activation of the variable $a$, to satisfy the Second Law (Dissipation Inequality) at every instant, while solving for the plastic strain $p$ and the dissipated energy $s^2/2$ in a well-set manner. This correction is achieved by using only the minimum amount of the \emph{control} variable $a$ necessary to restore non-negative dissipation which, in this case, drives the dissipation to exactly zero. The resulting stress-strain relations exhibit an elastic gap, which emerges as a direct consequence of the plastic strain remaining constant while $a$ is active. 

The close resemblance between the numerical and closed-form solutions is a consequence of the small initial value of $\bar{s}$ used. As mentioned in Section \ref{sec:results_sec}, the initial base state value acts as a selection parameter among the infinite family of solutions of the primal system, and larger initial values of $\bar{s}$ converged instead to different solutions, compared to the analytical case. The numerical solutions obtained with the gradient flow cum Newton-Raphson scheme, for both $m=1$ and $m=0.1$ cases, closely reproduce the analytical answer, achieving errors below $0.1\%$ throughout the domain, and only locally increasing where the modulating function $g$ rapidly transitions between positive and negative values.

These results provide a first numerical confirmation that the dual variational principle proposed in \cite{acharya2025second}, and the scheme proposed in \cite{AV_nash} are capable of enforcing the Second Law of Thermodynamics in a problem where the Second Law may not be satisfied in some process, using the constitutive model deemed adequate for the purpose. Short of improving the physics of the specification, say due to the absence of more reliable information and to avoid further ad-hoc phenomenology, one remedy involves the approach adopted herein. In the specific problem considered, the scheme results in the occurrence of `elastic gaps' in stress-strain response, reminiscent of behavior in some strain-gradient plasticity theories \cite{FHW}. However, the main applications of the technique are expected to be where the restrictions arising from the Second Law on constitutive functions become extremely complex due to the intricate nature of the mechanics of the model, or in the coupling of established models for disparate phenomena, or in implementing postulates like maximum dissipation  as a selection criterion in the analysis of nonlinear transport phenomena.

\printbibliography
\end{document}